\documentclass[10pt,leqno]{amsart}

\usepackage{graphicx}
\usepackage{indentfirst,csquotes}
\usepackage{amssymb,amsthm,amsmath}
\usepackage{xcolor,paralist,hyperref,titlesec,fancyhdr,etoolbox}
\usepackage{lipsum}
\usepackage{epsfig,amssymb,amsmath}
\usepackage{cite}
\usepackage{amsmath,amssymb,amsfonts}
\usepackage{algorithmic}
\usepackage{graphicx}
\usepackage{textcomp}
\usepackage{enumitem} % <-- needed for [label=...] in enumerate
\usepackage{xcolor}
\usepackage[utf8]{inputenc}
\usepackage{url}
\usepackage{booktabs}
\usepackage{caption}
\usepackage{hyperref}
\usepackage{pifont}
\usepackage{longtable}
\usepackage{adjustbox}
\usepackage{pdflscape}
\usepackage[table]{xcolor}
\usepackage{xcolor}
\usepackage{orcidlink}
\usepackage{xcolor}
\usepackage{listings}
\usepackage{caption}
\usepackage{float}
\usepackage{multirow}
\usepackage{tikz}
\usepackage[utf8]{inputenc}
\usepackage{csquotes}
\usepackage{amsaddr}
\usetikzlibrary{arrows.meta,positioning,shadows.blur}

\hypersetup{colorlinks=true, linkcolor=black, filecolor=black, urlcolor=black}

\titleformat{\section}[block]{\normalfont\Large\bfseries\centering}{\thesection}{1em}{}
\titlespacing*{\section}{0pt}{1ex}{0.5ex}

\makeatletter
\providecommand{\@secnumpunct}{\quad} % space (or use {. } for a dot)
\makeatother

\definecolor{remix-bg}{HTML}{1E1E1E}
\definecolor{remix-keyword}{HTML}{C586C0}
\definecolor{remix-type}{HTML}{4EC9B0}
\definecolor{remix-variable}{HTML}{9CDCFE}
\definecolor{remix-string}{HTML}{CE9178}
\definecolor{remix-comment}{HTML}{6A9955}
\definecolor{remix-function}{HTML}{DCDCAA}
\definecolor{remix-text}{HTML}{D4D4D4}

\lstdefinelanguage{Solidity}{
  morekeywords=[1]{contract, function, public, private, external, internal, view, pure, payable, return, returns, modifier, if, else, for, while, break, continue, revert, require, emit, event, new, try, catch, assert},
  morekeywords=[2]{uint, int, bool, string, address, bytes32, bytes, mapping, memory, storage, this, true, false},
  morekeywords=[3]{msg, sender, value, block, timestamp, now},
  sensitive=true,
  morecomment=[l]{//},
  morecomment=[s]{/*}{*/},
  morestring=[b]"
}

\begin{document}

\title{Attack-Centric by Design: A Program-Structure Taxonomy of Smart Contract Vulnerabilities}

\author{
Parsa Hedayatnia\,\orcidlink{0009-0008-0379-8165}\textsuperscript{1},
Tina Tavakkoli\,\orcidlink{0009-0001-8766-6897}\textsuperscript{1},
Hadi Amini\,\orcidlink{0009-0006-6848-4904}\textsuperscript{1},\\
Mohammad Allahbakhsh\,\orcidlink{0000-0002-2861-7745}\textsuperscript{1,*},
Haleh Amintoosi\,\orcidlink{0000-0002-1447-8086}\textsuperscript{1}
}

% ---- SINGLE SHARED AFFILIATION ----
\address{\textsuperscript{1}Computer Engineering Department, Faculty of Engineering, Ferdowsi University of Mashhad (FUM), Mashhad, Iran}

\email{\parbox{\textwidth}{
\texttt{parsa.hedayatnia@alumni.um.ac.ir},\,
\texttt{tina.tavakkoli@mail.um.ac.ir},\,
\texttt{hadi.amini@mail.um.ac.ir},\,
\texttt{allahbakhsh@um.ac.ir},\,
\texttt{amintoosi@um.ac.ir}
}}
\thanks{\textsuperscript{*}Corresponding author: \href{mailto:allahbakhsh@um.ac.ir}{allahbakhsh@um.ac.ir}}

\date{\today}

% ---- Abstract MUST precede \maketitle in amsart ----
% =================== ABSTRACT (replace your empty abstract) ===================

\maketitle

% (Optional) MSC footnote without breaking anchors
\renewcommand{\thefootnote}{\fnsymbol{footnote}}
\footnotetext[1]{MSC2020: Primary 00A05, Secondary 00A66.}
\renewcommand{\thefootnote}{\arabic{footnote}}
\begin{abstract}
Smart contracts concentrate high-value assets and complex logic in small, immutable programs, where even minor bugs can cause major losses. Existing taxonomies and tools remain fragmented—organized around symptoms such as reentrancy rather than structural causes. This paper introduces an \emph{attack-centric, program-structure} taxonomy that unifies Solidity vulnerabilities into eight root-cause families covering control flow, external calls, state integrity, arithmetic safety, environmental dependencies, access control, input validation, and cross-domain protocol assumptions. Each family is illustrated through concise Solidity examples, exploit mechanics, and mitigations, and linked to the detection signals observable by static, dynamic, and learning-based tools. We further cross-map legacy datasets (SmartBugs, SolidiFI) to this taxonomy to reveal label drift and coverage gaps. The taxonomy provides a consistent vocabulary and practical checklist that enable more interpretable detection, reproducible audits, and structured security education for both researchers and practitioners.

\end{abstract}
\begingroup
\small
\noindent\textbf{Keywords—}
Smart Contracts, Blockchain Security, Vulnerability Taxonomy, Solidity Security, Program Structure, Attack Surface, Root-Cause Analysis, Attack-Centric Classification.
\par
\endgroup
\medskip

\section{Introduction}
Smart contracts encode financial logic and governance rules that execute autonomously on blockchains. Their immutability, composability, and ubiquitous deployment elevate the security impact of defects: a single vulnerability can be replicated across thousands of instances, exploited permissionlessly, and amplified through inter-contract interactions. Over the last decade, the community has catalogued numerous failure modes and high-profile exploits, beginning with The Decentralized autonomous organization (DAO) and continuing through reentrancy variants, randomness manipulation, access-control bypasses, and proxy/upgradeability pitfalls. Early systematization by Atzei et al.~\cite{atzei2017survey} established a layered view (Solidity, Ethereum Virtual Machine (EVM), blockchain) that has since been refined by broad surveys and mappings~\cite{chu2023survey, vidal2024vulnerability, jiao2024survey, kezadri2025ethereum, zhu2024survey, iuliano2024smart}.

%\paragraph{Motivation and scope.}
This article is a \emph{review focused on Solidity vulnerabilities}. Rather than centering on any single detection technique, we provide a consolidated, program-structure–oriented taxonomy of attack surfaces and show how concrete exploits arise from (i) control-flow interaction (e.g., reentrancy, transaction ordering dependence), (ii) external calls and exception handling, (iii) state/storage integrity (e.g., unsafe \texttt{delegatecall}), (iv) arithmetic and type safety, (v) blockchain-environment dependency (e.g., timestamp and on-chain randomness), (vi) access control/authentication (e.g., \texttt{tx.origin} misuse), (vii) Application Binary Interface (ABI)/input validation (e.g., short-address), and (viii) network/protocol issues. For each family, we relate historical exploits to formal root causes (data flow, control flow, storage aliasing, environment assumptions), bridging legacy names with modern Smart Contract Weakness Classification (SWC)-style categories and recent taxonomies.

%\paragraph{Why another taxonomy now?}
Three trends justify a fresh, attack-centric consolidation. \emph{First}, detection and assurance techniques have diversified: precise Control Flow Graph (CFG) recovery for bytecode~\cite{pasqua2023enhancing}, expert-rule/static analyzers~\cite{liu2023smart, wang2025contractsentry}, dynamic/fuzzing ecosystems and injected benchmarks~\cite{ferreira2020smartbugs, ghaleb2020effective}, and multiple deep-learning lines—Abstract Syntax Tree (AST)/CFG/Graph neural network (GNN) and transformer models~\cite{tang2023deep, cai2023combine, gong2023scgformer, zhang2023svscanner, li2023smart, yuan2023optimizing, hwang2022codenet, huang2022smart}. \emph{Second}, the attack surface has shifted with upgradeable proxies, cross-contract patterns, and “long” contracts, pushing research toward multimodal fusion and efficiency~\cite{deng2023smart, colin2024integrated, fan2025small, lin2025lightweight}. \emph{Third}, large language models (LLMs) and LLM-enhanced frameworks are being actively explored for triage, explanation, and hybrid pipelines~\cite{boi2024smart, chen2025chatgpt, ding2025smartguard}.

\paragraph{Our contributions are as follows:}
\begin{compactitem}
  \item A unified, attack-centric taxonomy aligned to program-structure root causes, reconciling legacy names with modern categories and mitigations~\cite{atzei2017survey, chu2023survey, vidal2024vulnerability, jiao2024survey, kezadri2025ethereum, iuliano2024smart}.
  \item A detector-signal view for each family (what static, dynamic, and learning-based tools must observe) and code-backed exemplars~\cite{ferreira2020smartbugs, ghaleb2020effective, pasqua2023enhancing, cai2023combine, gong2023scgformer, zhang2023svscanner, boi2024smart, chen2025chatgpt, ding2025smartguard, otoni2023solicitous}.
  \item A cross-mapping of benchmark labels (SmartBugs, SolidiFI, and derivatives) to reduce ambiguity and surface coverage gaps and sources of label drift~\cite{ferreira2020smartbugs, ghaleb2020effective, iuliano2024smart}.
  \item We highlight open challenges: (i) precise inter-contract reasoning at bytecode level~\cite{pasqua2023enhancing}, (ii) explainability and false-positive control for AI models~\cite{vidal2024vulnerability, zhu2024survey, wu2024review}, and (iii) scalable formal verification for real-world patterns (loops, calls, upgradeability)~\cite{otoni2023solicitous, garfatta2021survey, olivieri2024software}.
  \item All taxonomies, code listings, and mappings are designed for reuse in benchmarking, education, and the development of future detection tools.
\end{compactitem}

While numerous works focus on \emph{how} to detect vulnerabilities—GNNs with expert priors~\cite{liu2021combining}, hierarchical attention~\cite{ma2023hgat}, CFG+Transformer~\cite{gong2023scgformer}, multimodal/entropy fusion~\cite{yuan2023optimizing, li2023smart}, BERT variants and semi-supervision~\cite{sun2023assbert}, and practical tool integrations~\cite{colin2024integrated, wang2025contractsentry}—our review centers on \emph{what gets attacked and why}. We reference detection/verification literature only to ground each attack family in concrete mitigations and analysis affordances.

% ========== FIXED "PAPER ORGANIZATION" PARAGRAPH IN INTRODUCTION ==========
\paragraph{\textbf{Paper organization.}\\}
Section~\ref{sec:related} reviews related taxonomies, vulnerability detectors, and benchmark datasets. 
Section~\ref{sec:taxonomy} introduces our unified vulnerability taxonomy grounded in eight program-structure root causes.
Section ~\ref{sec:vulnerabilities} presents illustrative vulnerability examples with annotated code, exploitation mechanics, and mitigation guidance.
Section~\ref{sec:discussion} discusses implications for benchmarking, explainability, and dataset relabeling.
section~\ref{sec:legacy} maps legacy/historical attack terminology to the unified taxonomy for consistent reporting. 
Section~\ref{sec:conclusion} concludes and outlines future work.

\section{Related literature}\label{sec:related}
Understanding and detecting smart contract vulnerabilities has spurred a rich and growing body of research, spanning taxonomic surveys, detection tools, evaluation benchmarks, and formal verification approaches. This section synthesizes the most influential studies across these themes to contextualize our taxonomy and highlight unresolved challenges. Whereas prior work often organizes vulnerabilities by language features, detection heuristics, or incident outcomes, our contribution is distinguished by an attack-centric taxonomy rooted in program-structure dimensions. This structure helps reconcile historical naming inconsistencies and aligns detection signals, tool coverage, and dataset design under a unified analytical framework. We group the related literature into six key strands: foundational taxonomies, evaluation benchmarks, program analysis, machine learning, environment-based attacks, and LLM-integrated pipelines.

\noindent\textbf{Foundational taxonomies and surveys.}
The SoK by Atzei et al.~\cite{atzei2017survey} introduced a layered taxonomy that remains a reference point for classifying Solidity/EVM/blockchain-level flaws. Recent surveys expand both breadth and depth: Chu et al.~\cite{chu2023survey} organize vulnerabilities alongside data sources and repair; Vidal et al.~\cite{vidal2024vulnerability} provide a systematic literature review emphasizing technique coverage and evaluation gaps; Jiao et al.~\cite{jiao2024survey} synthesize attacks and detection for Ethereum specifically; Kezadri Hamiaz \& Driss~\cite{kezadri2025ethereum} catalog tools and challenges across verification techniques; Zhu et al.~\cite{zhu2024survey} compare formal, fuzzing, Machine Learning (ML), and program-analysis pipelines; and Iuliano \& Di~Nucci~\cite{iuliano2024smart} review vulnerabilities, tools, and benchmarks up to 2024, noting deficiencies in standard datasets and labeling.

\medskip
\noindent\textbf{Datasets, benchmarks, and evaluation.}
SmartBugs~\cite{ferreira2020smartbugs} curated real and synthetic vulnerable contracts with a uniform runner for tool evaluation; SolidiFI~\cite{ghaleb2020effective} introduced systematic bug injection for controlled benchmarking. Subsequent works reuse or extend these corpora (e.g., \cite{colin2024integrated, li2023smart, yuan2023optimizing}) while highlighting threats to validity such as label noise and version skew~\cite{iuliano2024smart}. Precise bytecode-level CFG reconstruction~\cite{pasqua2023enhancing} improves static analyses and reduces spurious alarms for flow-sensitive categories (reentrancy, call-before-effects).

\medskip
\noindent\textbf{Static/dynamic program analysis.}
Representative static analyzers and hybrids span expert rules and symbolic reasoning~\cite{liu2023smart, wang2025contractsentry}, with ongoing interest in exception handling, external calls, and proxy/upgrade patterns. Dynamic techniques and fuzzing ecosystems are commonly evaluated on SmartBugs/SolidiFI runners~\cite{ferreira2020smartbugs, ghaleb2020effective}. Formal verification advances include modular reasoning and invariant synthesis for realistic Solidity patterns~\cite{otoni2023solicitous}, surveys of verification approaches~\cite{garfatta2021survey}, and cross-layer verification challenges~\cite{olivieri2024software}.

\medskip
\noindent\textbf{Learning-based detection.}
Deep learning directions capture syntax, structure, and semantics through AST/CFG graphs, token sequences, and opcode views: CNN/RNN hybrids~\cite{hwang2022codenet, zhang2022cbgru}, GNNs with expert priors~\cite{liu2021combining, cai2023combine, ma2023hgat}, transformer and attention-based models~\cite{gong2023scgformer, zhang2023svscanner}, and multimodal/entropy or decision fusion~\cite{tang2023deep, yuan2023optimizing, li2023smart, deng2023smart, colin2024integrated}. Semi-supervised and active-learning BERT variants~\cite{sun2023assbert} address label scarcity; recent work targets long contracts and few-shot regimes via feature fusion~\cite{fan2025small, lin2025lightweight}. Surveys focused on Deep Learning tooling report strengths and common failure modes~\cite{wu2024review}.

\medskip
\noindent\textbf{Environment-driven attacks.}
Randomness misuse and miner influence on timestamps/blocks have been systematically analyzed by Qian et al.~\cite{qian2023demystifying}, providing a taxonomy and detectors that complement our environment-dependency family. Access-control pitfalls such as \texttt{tx.origin} misuse and suicidal contracts persist in both rule-based and learning datasets~\cite{zhang2022cbgru, chu2023survey, colin2024integrated}.

\medskip
\noindent\textbf{LLMs and hybrid pipelines.}
Recent studies discuss where LLMs help and where they fail in vulnerability detection: position papers and evaluations~\cite{boi2024smart, chen2025chatgpt} and LLM-enhanced frameworks such as SmartGuard~\cite{ding2025smartguard}. These works motivate explainable pipelines that pair LLM reasoning with static/dynamic evidence---especially for composite attacks and cross-contract behaviors.

\medskip
In contrast to the above, our article keeps the \emph{attack} as the central unit and uses a program-structure–oriented taxonomy to align historic exploits, modern terminology, datasets, and detection/verification evidence. This framing simplifies mapping legacy names to unified families and clarifies which analysis capabilities are necessary to prevent each class of exploit.

\section{Taxonomy of Smart Contract Vulnerabilities}\label{sec:taxonomy}

As Ethereum smart contracts grow more complex and handle greater value, security risks stemming from subtle design flaws have become increasingly critical. Despite the availability of multiple vulnerability lists and static analysis tools, inconsistencies in naming, classification, and coverage persist. Some taxonomies group issues by symptoms (e.g., ``reentrancy''), while others follow tool-driven patterns. However, these approaches often lack a consistent basis for identifying root causes or guiding detection and mitigation.

To address this, we propose a unified taxonomy grounded in \emph{program structure}---how contracts process control, handle data, validate inputs, and interact with their environment. By focusing on the structural source of each vulnerability, our taxonomy helps bridge the gap between diverse historical terminology and modern detection techniques, making it easier to reason about, detect, and prevent vulnerabilities.

\subsection*{The Eight Root-Cause Families}
We classify vulnerabilities into eight families (visualized in Figure~\ref{fig:taxonomy-pdf}):

\begin{samepage}
\noindent\begin{minipage}{0.48\textwidth}
\begin{enumerate}[leftmargin=*,labelsep=0.3em]
  \item \textbf{Control-Flow Interaction} (e.g., reentrancy)
  \item \textbf{External Calls \& Error Handling}
  \item \textbf{State/Storage Integrity} (e.g., unsafe \texttt{delegatecall})
  \item \textbf{Arithmetic \& Type Safety}
\end{enumerate}
\end{minipage}
\hfill
\begin{minipage}{0.48\textwidth}
\begin{enumerate}[leftmargin=*,labelsep=0.3em]
  \setcounter{enumi}{4}
  \item \textbf{Environment Dependency} (e.g., timestamp)
  \item \textbf{Access Control}
  \item \textbf{ABI/Input Validation}
  \item \textbf{Network/Protocol Layer} (e.g., replay attacks)
\end{enumerate}
\end{minipage}
\end{samepage}

\vspace{1em}
\subsection*{Design Principles}
Each family is designed to be:
\begin{itemize}
  \item \textbf{Root-cause aligned}: Groups flaws by shared structural origin.
  \item \textbf{Tool-compatible}: Maps directly to signals used by static, dynamic, and ML detectors.
  \item \textbf{Education-friendly}: Paired with minimal code, exploit trace, and fix.
\end{itemize}

\subsection*{Placement Criteria}
A vulnerability is assigned to the family that \emph{most directly} explains its exploit and suggests its fix. We use six lightweight criteria:

\begin{enumerate}[label=(\alph*)]
  \item \emph{Control flow}: Does the issue involve execution order? (e.g., reentrancy)
  \item \emph{Storage access}: Is state aliased or corrupted? (e.g., \texttt{delegatecall})
  \item \emph{Environment}: Does blockchain state affect logic? (e.g., \texttt{block.timestamp})
  \item \emph{Identity}: Are permissions checked correctly? (e.g., \texttt{tx.origin})
  \item \emph{Input/ABI}: Is decoding ambiguous? (e.g., packed encoding)
  \item \emph{Cross-domain}: Does it span chains or protocols? (e.g., signature replay)
\end{enumerate}

\begin{figure}[h]
  \centering
  \includegraphics[width=0.95\textwidth]{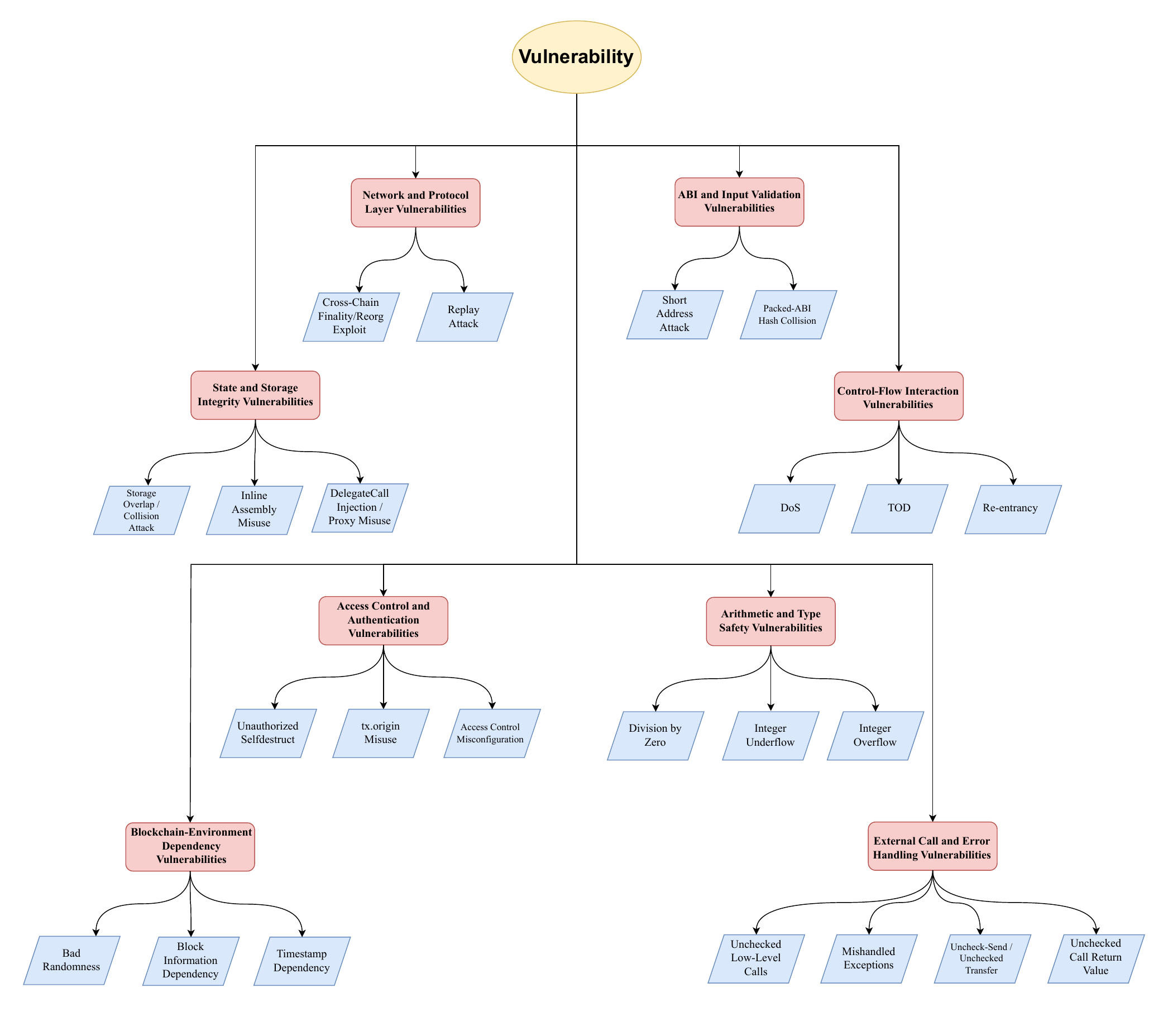}
  \caption{Program-structure taxonomy: eight families, common attacks, and canonical defenses.}
  \label{fig:taxonomy-pdf}
\end{figure}

Figure~\ref{fig:taxonomy-pdf} organizes these criteria clockwise by reasoning complexity, from local arithmetic to cross-chain assumptions.

\subsection*{Impact and Utility}

This taxonomy enables more effective detection, auditing, and classification in both research and practice. It helps:
\begin{itemize}
  \item Unify terminology across legacy datasets, reports, and tools;
  \item Clarify the intent of each detection technique by aligning it with structural causes;
  \item Guide developers toward better mitigation by identifying consistent patterns;
  \item Provide a reliable foundation for dataset labeling, evaluation, and benchmarking.
\end{itemize}

Instead of treating vulnerabilities as isolated exceptions, this classification reveals their shared structural more robust and systematic security analysis across the entire lifecycle of smart contract development.

\section{Smart Contract Vulnerabilities }\label{sec:vulnerabilities}
Smart contracts, by design, encode and enforce rules autonomously on blockchain platforms without centralized intervention. However, this autonomy makes them especially vulnerable to logic errors, cryptoeconomic flaws, and environment-specific assumptions. In this section, we present a structured walkthrough of vulnerability classes as categorized in our taxonomy (Figure~\ref{fig:taxonomy-pdf}). Each vulnerability is analyzed in terms of its definition, technical explanation, real-world exploitation scenarios, and mitigation recommendations. The goal is to provide developers, auditors, and researchers with actionable insight into the mechanics behind recurring smart contract flaws. To enhance clarity, we group vulnerabilities into eight root-cause families, such as control-flow mismanagement, access control misconfiguration, and reliance on unpredictable blockchain state. For each entry, we also link representative code listings and note how legacy terminology aligns with the modern classification. This consolidated view enables consistent understanding and sets a foundation for tool evaluation, security auditing, and future research in smart contract safety.

\subsection{Control-Flow Interaction Vulnerabilities}
\paragraph{
This family covers flaws that break the intended execution order within or across contracts. 
They typically arise when developers assume atomic behavior in a non-atomic environment, allowing adversaries to interleave or reorder transactions. 
As shown in Figure~\ref{fig:taxonomy-pdf}, these vulnerabilities form the first layer of the taxonomy, where execution flow integrity governs all subsequent state and logic correctness.}

\subsubsection{Re-entrancy}\label{vuln:reentrancy}
\paragraph{\textbf{Definition:}}
Reentrancy is a vulnerability that arises when an external contract invokes a function in the victim contract before the previous execution completes. If the internal state of the contract is not updated before the external call, the attacker can exploit the recursive flow to repeatedly trigger sensitive functions such as withdrawals~\cite{ferreira2020smartbugs, gong2023scgformer}.

\begin{lstlisting}[caption={Reentrancy Vulnerability with Fake Guard}, label={lst:reentrancy}]
contract Reentrancy {
    mapping(address => uint256) public balances;
    bool locked = false;

    function deposit() public payable {
        balances[msg.sender] += msg.value;
    }

    function withdraw() public {
        require(!locked, "Reentrant call");  // Vulnerability: Reentrancy guard check exists
        locked = true;

        (bool success, ) = msg.sender.call{value: balances[msg.sender]}("");  // Vulnerability: External call before state change
        require(success, "Transfer failed");

        balances[msg.sender] = 0;

        // Fake unlock - conditional and unsafe
        if (balances[msg.sender] == 0) {
            locked = false;  // Vulnerability: Lock released conditionally
        }
    }
}
\end{lstlisting}

\paragraph{\textbf{Explanation of Vulnerability:}}

This contract implements a \texttt{locked} flag as a reentrancy guard; however, its logic is flawed:

\begin{itemize}
    \item \textbf{Line 6 – }
     Ether is transferred to \texttt{msg.sender} using a low-level \texttt{call} before the balance is cleared.
    \item \textbf{Line 10 and Line 19 – } 
     Although a reentrancy guard is introduced with the \texttt{locked} flag, its reset depends on the condition \texttt{balances[msg.sender] == 0}, which is attacker-controlled. During reentrant execution, this condition may never be satisfied, allowing multiple recursive invocations.
\end{itemize}

\paragraph{\textbf{Why This Is a Vulnerability:}} 
\begin{itemize}
  \item \textbf{Violation of CEI discipline:} The contract performs an external call before updating its internal state, breaking the Checks–Effects–Interactions (CEI) pattern and exposing inconsistent state during execution.  
  \item \textbf{Synchronous execution model:} Because Ethereum executes external calls synchronously, the callee can re-enter the caller’s function before the first execution completes, allowing repeated access to mutable state variables.  
  \item \textbf{Deferred state updates:} Sensitive values such as balances are updated only after the external call, allowing attackers to reuse stale data from previous invocations.  
  \item \textbf{Weak or conditional reentrancy guard:} Guards that depend on attacker-controlled conditions or are reset inconsistently (e.g., conditional unlocks) fail to prevent recursive entry.  
  \item \textbf{Post-call invariant enforcement:} Logic that enforces key invariants (e.g., resetting balances) after the external interaction is vulnerable to manipulation during reentrant calls.  
\end{itemize}

As a result, internal accounting and control flow can become inconsistent, enabling unauthorized withdrawals, recursive fund extraction, or logical bypasses during reentrant invocations.

\paragraph{\textbf{Exploitation Scenario:}}
\begin{itemize}
    \item An attacker deploys a contract with a fallback function that calls \texttt{withdraw()} when Ether is received.
    \item Upon the first \texttt{withdraw()} call, Line 10 triggers the fallback function before Line 12 clears the balance.
    \item The attacker recursively invokes \texttt{withdraw()} multiple times and receives Ether in each iteration.
    \item Since the \texttt{locked = false} statement is placed conditionally and may never execute, the reentrancy guard fails entirely.
\end{itemize}

\paragraph{\textbf{Why the Guard is Ineffective (Fake Guard):}}
\begin{itemize}
    \item A valid reentrancy guard must be enforced at the beginning and released deterministically at the end of execution.
    \item In this example, the unlock logic (Line 14) is based on a balance check that may not hold if recursive calls interfere, rendering the lock useless.
\end{itemize}

\paragraph{\textbf{Mitigation Recommendations:}}
\begin{itemize}
    \item Use the checks-effects-interactions pattern by placing the state update (\texttt{balances[msg.sender] = 0}) before the external \texttt{call}.
    \item Employ well-audited libraries such as OpenZeppelin’s \texttt{nonReentrant} modifier.
    \item Ensure that any guard variables such as \texttt{locked} are always reset unconditionally at the end of execution.
\end{itemize}

\subsubsection{Transaction Order Dependency (TOD) / Frontrunning}\label{vuln:txorder}
\paragraph{\textbf{Definition:}}
Transaction Order Dependence (TOD) refers to a vulnerability where the correctness or fairness of contract logic depends on the ordering of transactions in a block. Since miners choose which transactions to include and in what order, adversaries can exploit this by front-running other users' transactions~\cite{ghaleb2020effective, deng2023smart}.

\begin{lstlisting}[caption={Transaction Ordering Dependence Vulnerability}, label={lst:txorder}]
contract TxOrderDependence {
    mapping(address => uint256) public balances;

    function donate() public payable {
        balances[msg.sender] += msg.value;
    }

    function withdraw() public {
        // Vulnerability: Transaction ordering dependence in withdrawal
        require(balances[msg.sender] > 0);
        msg.sender.transfer(balances[msg.sender]);
        balances[msg.sender] = 0;
    }

    function commitReveal(bytes32 hash) public {
        // Vulnerability: Unvalidated commit input, lacks sender binding
        storedHash = hash;
    }

    function reveal(uint secret) public {
        // Vulnerability: No link to original sender in commit-reveal scheme
        require(keccak256(abi.encodePacked(secret)) == storedHash);
        reward[msg.sender] = secret;
    }

    uint public nonce;
    function increment(uint input) public {
        // Warning: nonce pattern can be abused without stronger validation
        require(input == nonce);
        nonce++;
    }
}
\end{lstlisting}

\paragraph{\textbf{Explanation of Vulnerability:}}

\begin{itemize}
    \item \textbf{Line 8 – Withdrawal Based on Mutable State }  
    A user’s balance is checked and transferred in the same function, making it susceptible to frontrunning before the balance is reset.

    \item \textbf{Line 15 – Missing Binding in Commit Phase }  
    The stored hash is global, not tied to a sender. Any caller can overwrite it, introducing a race condition.

    \item \textbf{Line 20 – Reveal Phase Unbound to Committer }   
    The reveal logic compares input to the shared hash but does not confirm the revealing party is the one who committed. This is exploitable by frontrunners.

    \item \textbf{Line 29 – Weak Nonce Usage }  
    The nonce is updated based on a public check and can be predicted or manipulated if reused or not protected contextually.
\end{itemize}

\paragraph{\textbf{Why This Is a Vulnerability:}} 
\begin{itemize}
  \item \textbf{Dependence on transaction sequencing:}  
  The correctness or fairness of the contract logic depends on the order in which transactions are included in a block. Since miners or validators can freely reorder transactions, execution outcomes become nondeterministic.

  \item \textbf{Mempool visibility and frontrunning:}  
  Pending transactions are publicly visible in the mempool. Adversaries can observe these and submit similar transactions with higher gas fees to be processed first, gaining unfair advantages.

  \item \textbf{Shared mutable state:}  
  Functions such as \texttt{withdraw()} rely on global state variables (\texttt{balances}) that multiple users can modify concurrently. This shared state enables race conditions when competing transactions are processed.

  \item \textbf{Lack of sender binding in commit–reveal patterns:}  
  Commit–reveal schemes that do not associate the commitment with a specific sender (e.g., \texttt{storedHash}) allow attackers to overwrite or reveal values on behalf of others.

  \item \textbf{Predictable or weak nonce usage:}  
  Publicly accessible counters or nonce variables can be anticipated or reused by attackers, allowing them to preempt legitimate operations or bypass intended sequencing constraints.
\end{itemize}
Such dependence on nondeterministic transaction ordering enables adversaries to front-run withdrawals, overwrite commits, or exploit race conditions—compromising fairness, consistency, and contract integrity.

\paragraph{\textbf{Exploitation Scenario:}}

\begin{itemize}
    \item A user submits a \texttt{withdraw()} call. An attacker sees it in the mempool and submits a higher-fee withdrawal that empties the balance first.
    \item During the \texttt{commitReveal()} phase, an attacker overwrites the committed hash. In the next block, they reveal their precomputed secret, stealing the reward.
    \item The \texttt{increment()} function can be replayed or bypassed if the nonce logic is used across contexts.
\end{itemize}

\paragraph{\textbf{Mitigation Recommendations:}}

\begin{itemize}
    \item Use pull-based withdrawal patterns with lock periods or commit delays to defend against frontrunning.
    \item Always bind commit phases to sender addresses: \texttt{commit[msg.sender] = hash}.
    \item Add timeouts, nonce expiry, or salt-based nonces to secure the commit-reveal pattern.
    \item Strengthen nonce control with role checks or context-aware validation.
\end{itemize}

\subsubsection{Denial of Service (DoS)}\label{vuln:dos}
\paragraph{\textbf{Definition:}}
Denial-of-Service (DoS) with revert occurs when smart contracts expose logic that can be intentionally halted or forced to fail, such as looping over external input or relying on untrusted receivers~\cite{pasqua2023enhancing}.\\
\begin{lstlisting}[caption={Denial-of-Service (DoS) Vulnerability }, label={lst:Dos}]
mapping(address => uint) public balances;
address[] public users;

function processPayments(uint iterations) public {
    for (uint i = 0; i < iterations; i++) {  
        // Vulnerability: loop controlled by external argument (DoS risk)
        
        if (balances[users[i]] == 0) {
            revert("Zero balance");  
            // Vulnerability: revert inside loop can halt all processing (DoS)
        }

        (bool success, ) = users[i].call{value: balances[users[i]]}("");  
        // Vulnerability: untrusted external call in loop (DoS and reentrancy risk)
        // Vulnerability: low-level call with result unchecked before state update

        balances[users[i]] = 0;
    }
}
\end{lstlisting}

\paragraph{\textbf{Explanation of Vulnerabilities:}}
\begin{itemize}
    \item \textbf{Line 5 – Argument-Controlled Loop } 
    The loop count is fully controlled by the caller through the \texttt{iterations} parameter. This allows gas griefing and unaudited unbounded loops.

    \item \textbf{Line 9 – Revert Inside Loop} 
    A revert due to a user having zero balance will break the entire function, stopping all further payments.

    \item \textbf{Line 13 – Untrusted External Call in Loop }
    Calling untrusted addresses inside a loop exposes the function to reentrancy, fallback abuse, or gas griefing by external contracts.

    \item \textbf{Line 13 – Unchecked Result of \texttt{call} }
    The return value of \texttt{call} is not verified. Failed Ether transfers will go unnoticed, and funds may remain locked or lost.
\end{itemize}

\paragraph{\textbf{Why This Is a Vulnerability:}}
\begin{itemize}
    \item Allows a single malicious or misconfigured user to halt or block contract execution.
    \item Violates principles of fault isolation and resilience.
    \item Enables gas exhaustion attacks or silent failures that break contract assumptions.
\end{itemize}

\paragraph{\textbf{Exploitation Scenario:}}
\begin{itemize}
    \item A malicious user sets \texttt{iterations} to a high number, causing gas exhaustion.
    \item A zero-balance user causes revert in the loop, blocking payments to others.
    \item A contract with a gas-heavy fallback function is inserted into \texttt{users}, making the call fail or waste gas.
\end{itemize}

\paragraph{\textbf{Mitigation Recommendations:}}
\begin{itemize}
    \item Avoid processing logic in unbounded loops controlled by external input.
    \item Validate user input and avoid critical reverts inside loops.
    \item Use the pull-payment pattern: let users claim funds individually.
    \item Check all low-level call results and handle failures gracefully:
\begin{lstlisting}[language=Solidity]
(bool success, ) = users[i].call{value: amount}("");
require(success, "Transfer failed");
\end{lstlisting}
    \item Split long operations into batches using state tracking across transactions.
\end{itemize}

% ============================================================

\subsection{External Call and Error Handling Vulnerabilities}
\paragraph{This family captures faults where contracts mishandle results of external interactions. 
Low-level primitives such as \texttt{call}, \texttt{send}, and \texttt{delegatecall} return a boolean success flag instead of reverting automatically. 
Failing to inspect these outcomes or to isolate fallback behavior leads to silent execution failures, inconsistent state, or exploitable control flow.}

\subsubsection{Unchecked Call Return Value}\label{vuln:unchecked-call}
\paragraph{\textbf{Definition:}}  
An \emph{unchecked call return value} vulnerability occurs when a smart contract performs a low-level external call (such as \texttt{call}, \texttt{delegatecall}, \texttt{staticcall}, or legacy \texttt{send}) but fails to verify the returned boolean result.  
Because these primitives do not automatically revert on failure, ignoring the return value can lead to \emph{silent execution failure}, leaving the contract in an inconsistent state or enabling attackers to manipulate business logic~\cite{atzei2017survey, ferreira2020smartbugs, ghaleb2020effective, chu2023survey, he2023detection}.

\begin{lstlisting}[caption={Unchecked call return value Vulnerability}, label={lst:unchecked_call}]
pragma solidity ^0.8.0;

contract PaymentProcessor {
    mapping(address => uint256) public balances;
    event PaymentSent(address indexed from, address indexed to, uint256 amount);

    function deposit() external payable {
        balances[msg.sender] += msg.value;
    }

    function pay(address payable recipient, uint256 amount) external {
        require(balances[msg.sender] >= amount, "Insufficient balance");

        // Effect: balance deducted before external call
        balances[msg.sender] -= amount;

        // Vulnerability: ignoring return value of low-level call
        recipient.call{value: amount}("");

        // Continues as if transfer succeeded even if it failed
        emit PaymentSent(msg.sender, recipient, amount);
    }
}
\end{lstlisting}

\paragraph{\textbf{Explanation of the Vulnerability:}}
In Listing~\ref{lst:unchecked_call}, the vulnerable behavior resides primarily on Lines~17–18 and 15–21, where the contract performs a low-level call without verifying its outcome and continues execution under the false assumption of success.  

\begin{itemize}
  \item \textbf{Line 15 – Premature State Update } The sender’s balance is deducted \emph{before} performing the external call. This violates the Checks–Effects–Interactions (CEI) pattern by changing state before verifying external interaction success.  

  \item \textbf{Line 18 – Ignored Return Value } The expression \texttt{recipient.call\{value: amount\}("")} executes an external call but disregards the returned boolean (\texttt{success}). If the callee reverts or runs out of gas, the call silently fails.  

  \item \textbf{Line 21 – Misleading Event Emission } Despite potential failure of the Ether transfer, Line~21 emits \texttt{PaymentSent}, signaling success even though no funds left the contract. This introduces logical inconsistency between on-chain logs and the actual state.  
\end{itemize}

If the recipient’s fallback function reverts or consumes excessive gas, the transaction fails internally but the contract does not revert. The balance remains reduced while the event log reports a successful payment, resulting in accounting mismatches or permanent fund loss.  
This logical inconsistency can result in fund loss, denial of service, or accounting errors.  
Static analysis frameworks (e.g., SmartBugs, SolidiFI) often flag such patterns as part of the \textit{External Call and Error Handling} category~\cite{ferreira2020smartbugs, he2023detection}.

\paragraph{\textbf{Why This Is a Vulnerability:}}
\begin{itemize}
  \item \textbf{Silent failure:} If the external call fails but its return value is unchecked, the contract continues execution under the false assumption of success.  
  \item \textbf{Broken invariants:} Internal state (such as user balances) no longer matches the actual Ether distribution, creating accounting inconsistencies.  
  \item \textbf{Exploitable failures:} Attackers may deliberately cause reverts to exploit refund mechanisms or repeatedly trigger inconsistent states~\cite{ghaleb2020effective, chu2023survey}.
\end{itemize}

\paragraph{\textbf{Exploitation Scenario:}} 
\begin{itemize}
  \item \textbf{Malicious recipient contract:}  
  An attacker deploys a recipient contract whose fallback function always reverts or consumes excessive gas, ensuring that any Ether transfer to it will fail silently.
  \item \textbf{Ignored call result:}  
  When the vulnerable \texttt{PaymentProcessor} contract executes  
  \texttt{recipient.call\{value: amount\}("")} (Line~18), it fails to check the returned boolean value. The call reverts internally, but the contract proceeds as if successful.
  \item \textbf{State desynchronization:}  
  Because the sender’s balance is already deducted (Line~15) and no revert occurs, the contract’s internal accounting no longer reflects its actual Ether balance.
  \item \textbf{False success signal:}  
  The \texttt{PaymentSent} event (Line~21) is still emitted, misleading observers and off-chain systems into believing that payment succeeded.
  \item \textbf{Exploit chain reaction:}  
  The attacker can exploit this inconsistency by repeatedly triggering refunds, forcing compensatory logic, or leveraging mismatched balances to drain funds or lock assets.  
\end{itemize}
\paragraph{\textbf{Mitigation Recommendations:}}
\begin{itemize}
  \item \textbf{Always verify the return value:}
  \begin{lstlisting}[language=Solidity]
(bool success, ) = recipient.call{value: amount}("");
require(success, "Transfer failed");
  \end{lstlisting}
  \item \textbf{Adopt the pull-payment pattern:} Instead of pushing funds via external calls, record user balances and let recipients withdraw them manually through a controlled \texttt{withdraw()} function~\cite{ghaleb2020effective}.  
  \item \textbf{Use the Checks–Effects–Interactions (CEI) pattern:} Perform all validations and internal state updates before making any external calls, and guard against reentrancy.  
  \item \textbf{Use verified libraries:} Rely on well-tested frameworks like OpenZeppelin’s \texttt{PullPayment} or \texttt{PaymentSplitter} for handling Ether transfers safely.  
  \item \textbf{Handle potential call failures gracefully:} Include fallback logic, retry mechanisms, or event logs that record unsuccessful payments for manual review~\cite{ferreira2020smartbugs, he2023detection}.
\end{itemize}

\paragraph{\textbf{Notes:}}
Unchecked call return values are among the most frequent smart contract weaknesses detected by both static and dynamic analysis tools.  
Despite their apparent simplicity, they often indicate improper error-handling discipline and can be precursors to critical financial inconsistencies, especially in payment systems and token contracts.

\subsubsection{Uncheck-Send / Unchecked Transfer}\label{vuln:Uncheck-Send / Unchecked Transfer}
\paragraph{\textbf{Definition:}} 
An \emph{unchecked send} (or unchecked transfer) occurs when a contract transfers Ether to an external address using low-level primitives such as \texttt{call}, \texttt{send}, or \texttt{transfer} without properly checking the outcome or without constraining when and to whom transfers occur. This includes (a) ignoring the boolean success return from low-level calls, and (b) performing unconditional transfers (e.g., always sending a fixed amount to \texttt{msg.sender}) that may be abused by an attacker or cause unexpected state inconsistencies~\cite{ferreira2020smartbugs, ghaleb2020effective, he2023detection, chu2023survey}.

\begin{lstlisting}[caption={Unchecked transfers Vulnerability}, label={lst:unchk_send}]
pragma solidity >=0.4.21 <0.6.0;

contract DocumentSigner {
  function bug_unchk_send7() public payable {
      msg.sender.transfer(1 ether); 
      //Vulnerability: unconditional transfer without authorization or state update
  }

  mapping(bytes32 => string) public docs;

  function bug_unchk_send23() public payable {
      msg.sender.transfer(1 ether); 
      // [Line 10] Vulnerability: no access control or success check
  }

  mapping(bytes32 => address[]) public signers;

  function bug_unchk_send31() public payable {
      msg.sender.transfer(1 ether); 
      //Vulnerability: repetitive value transfer without validation or accounting
  }

  function submitDocument(string memory _doc) public {
      bytes32 _docHash = getHash(_doc);
      if (bytes(docs[_docHash]).length == 0) {
          docs[_docHash] = _doc;
          emit NewDocument(_docHash);
      }
      //Safe logic; unrelated to transfer vulnerability
  }

  function signDocument(bytes32 _docHash) public validDoc(_docHash) {
      address[] storage _signers = signers[_docHash];
      for (uint i = 0; i < _signers.length; i++) {
          if (_signers[i] == msg.sender) return;
      }
      _signers.push(msg.sender);
      //benign functionality; included for context
  }
}
\end{lstlisting}

\paragraph{\textbf{Explanation of the vulnerability:}} 
In Listing~\ref{lst:unchk_send}, several functions transfer Ether directly to the caller without verification or access control. 
\textbf{\\Line 4 -} The function \texttt{bug\_unchk\_send7()} performs an unconditional \texttt{msg.sender.transfer(1 ether)} call without checking authorization or success, allowing any user to withdraw Ether arbitrarily.  
\textbf{\\Line 11 -} The function \texttt{bug\_unchk\_send23()} repeats the same pattern, again omitting success handling and enabling silent failures if the transfer reverts due to gas or fallback execution.  
\textbf{\\Line 18 -} The function \texttt{bug\_unchk\_send31()} further illustrates unsafe code reuse. Each of these functions can be exploited independently to drain funds since no accounting or role validation is enforced.

\paragraph{\textbf{Why This Is a Vulnerability:}}
\begin{itemize}
  \item \textbf{Immediate fund leakage:} Any caller can repeatedly invoke these functions to drain Ether since transfers occur without accounting updates or restrictions.
  \item \textbf{Silent failure and broken invariants:} If a transfer fails (e.g., due to insufficient gas or receiver revert), the contract continues execution as though successful, leading to discrepancies between recorded and actual balances.
  \item \textbf{Gas and griefing exposure:} Older \texttt{transfer}/\texttt{send} semantics forward only 2300 gas. Attackers can exploit gas-heavy fallback functions to trigger failures or unexpected state behavior~\cite{ferreira2020smartbugs, chu2023survey}.
\end{itemize}

\paragraph{\textbf{Exploitation scenario:}} 
\begin{itemize}
  \item An attacker repeatedly calls any of the \texttt{bug\_unchk\_send*} functions to withdraw one ether per call until the contract balance reaches zero. No authentication or balance accounting prevents this drain.
  \item Alternatively, if a benign recipient has a gas-intensive fallback function, the transfer fails silently, leaving state inconsistent while the contract assumes success—opening room for double spends or accounting manipulation.
\end{itemize}

\paragraph{\textbf{Mitigation recommendations:}} 
\begin{itemize}
  \item \textbf{Avoid unconditional transfers:} Always link transfers to verified conditions (\texttt{onlyOwner}, \texttt{balances[msg.sender] > amount}) and perform updates before external calls.
  \item \textbf{Prefer pull-over-push payments:} Let users withdraw via a controlled \texttt{withdraw()} function rather than automatic transfers, minimizing reentrancy and gas griefing.
  \item \textbf{Check return values explicitly:}
\begin{lstlisting}[language=Solidity]
(bool success, ) = payable(recipient).call{value: amount}("");
require(success, "Transfer failed");
\end{lstlisting}
  \item \textbf{Use secure libraries:} Employ verified payment abstractions such as OpenZeppelin’s \texttt{PullPayment} or \texttt{PaymentSplitter}.
  \item \textbf{Enforce access control:} Add \texttt{onlyOwner} or role-based modifiers and rate limits to all Ether transfer functions.
  \item \textbf{Document expected failure behavior:} Handle benign failures gracefully (refunds, retries) and audit all unchecked send warnings flagged by Static Application Security Testing (SAST) tools.
\end{itemize}
\paragraph{\textbf{Notes:}}
\begin{itemize}
  \item \texttt{transfer()} and \texttt{send()} historically forwarded a fixed 2300 gas stipend; in modern environments using \texttt{call{value:...}("")} combined with an explicit `require(success)` is the recommended approach, but it must be paired with CEI and reentrancy protections~\cite{chu2023survey}.  
  \item Static analysis datasets (SmartBugs, SolidiFI) and recent detector evaluations frequently mark unchecked sends as a distinct family of defects because they are both common and high-impact; prioritize fixing them in audits and test harnesses~\cite{ferreira2020smartbugs, ghaleb2020effective}.
\end{itemize}

\subsubsection{Mishandled Exceptions}\label{vuln:mishandled}
\paragraph{\textbf{Definition:}}
Mishandled exceptions occur when errors such as failed external calls, invalid inputs, or logic exceptions are not properly checked or recovered from. This leads to silent failures or wasted gas and hinders reliable contract behavior~\cite{cai2023combine, ren2023smart}.

\begin{lstlisting}[caption={Mishandled Exceptions Vulnerability}, label={lst:mishandled}]
contract MishandledExceptions {
    function unsafeDivision(uint a, uint b) public pure returns (uint) {
        return a / b; // Vulnerability: division without zero check
    }

    function callWithoutRevert(address target) public {
        (bool success, ) = target.call("");
        if (!success) {
            // Vulnerability: external call failure ignored
        }
    }

    function withEmptyCatch() public {
        try this.unsafeDivision(10, 0) {
        } catch {
            // Vulnerability: empty catch block without handling logic
        }
    }

    function useAssert(uint a) public pure {
        assert(a != 0); // Vulnerability: assert used instead of require for input validation
    }
}
\end{lstlisting}
\paragraph{\textbf{Explanation of Vulnerability:}}
\begin{itemize}
    \item \textbf{Line 3 – Unsafe Division Without Check} \\
    There is no check that \texttt{b} is non-zero before performing division, risking a crash due to divide-by-zero.

    \item \textbf{Line 9 – Unhandled Call Failure} \\
    The call result is checked but ignored. There is no revert, log, or handling, leading to hidden failures.

    \item \textbf{Line 15 – Empty Catch Block} \\
    The catch block is syntactically present but lacks logic. This suppresses errors that could indicate major issues.

    \item \textbf{Line 21 – Misuse of \texttt{assert}} \\
    The \texttt{assert} statement is used for input validation, which is incorrect. Assertions are for developer invariants and consume all remaining gas on failure.
\end{itemize}

\paragraph{\textbf{Why This Is a Vulnerability:}}
\begin{itemize}
    \item Exceptions are part of critical control flow. Ignoring or misusing them makes contract behavior unreliable.
    \item Developers and auditors may not detect silent failures.
    \item Misuse of \texttt{assert} can crash the contract with high gas loss and no user-facing error.
\end{itemize}

\paragraph{\textbf{Exploitation Scenario:}}
\begin{itemize}
    \item An attacker could exploit silent failure to bypass logic intended to restrict behavior.
    \item Calls that appear to succeed may be quietly failing, leading to loss of funds or broken logic.
    \item Misused \texttt{assert()} can be weaponized to intentionally waste gas and disable contract functionality.
\end{itemize}

\paragraph{\textbf{Mitigation Recommendations:}}
\begin{itemize}
    \item Always perform zero checks before division.
    \item Revert explicitly after failed external calls:
    \begin{lstlisting}[language=Solidity]
(bool success, ) = target.call("");
require(success, "Call failed");
    \end{lstlisting}

    \item Avoid empty \texttt{catch} blocks. Add logs or safe fallbacks.
    \item Replace \texttt{assert()} with \texttt{require()} for all input validations.
\end{itemize}

\subsubsection{Unchecked Low-Level Calls}\label{vuln:Unchecked Low-Level Calls in Solidity}
\paragraph{\textbf{Definition:}}
Unchecked send occurs when a contract sends Ether or executes low-level calls using \texttt{call()}, \texttt{delegatecall()}, or \texttt{send()} without verifying whether the call succeeded. This may result in silent failure and unintended behavior, including loss of funds or unsafe execution paths~\cite{he2023detection}.
\begin{lstlisting}[caption={Unchecked Low-Level Calls Vulnerability}, label={lst:unchecked}]
contract UncheckedCalls {
    function withdraw(address payable user, uint amount) public {
        user.call{value: amount}("");
    }

    function sendEther(address payable recipient, uint amount) public {
        (bool success, ) = recipient.call{value: amount}("");
        // success is ignored, not checked
    }

    function unsafeDelegate(address target, bytes memory data) public {
        (bool ok, ) = target.delegatecall(data);
        if (ok) {
            // do something
        }
        // missing revert on failure
    }
}
\end{lstlisting}

\paragraph{\textbf{Explanation of Vulnerability:}}
\begin{itemize}
    \item \textbf{Line 3 – Unchecked \texttt{call} - } 
    Ether is sent using \texttt{user.call} without capturing or checking the result, potentially leading to untracked failures.

    \item \textbf{Line 7 – Call With Ignored Success Flag - } 
    Although the success flag is captured, it is never validated. This is functionally equivalent to ignoring the outcome.

    \item \textbf{Lines 11–13 – Silent \texttt{delegatecall} Failure - }
    The contract checks the result of the delegate call but fails to revert on failure. This allows execution to continue even if a critical subroutine fails.
\end{itemize}

\paragraph{\textbf{Why This Is a Vulnerability:}}
\begin{itemize}
    \item Solidity’s low-level calls return a success boolean instead of throwing an exception.
    \item Ignoring this boolean leads to logic that assumes success even when the call failed.
    \item This can break invariants, silently lose Ether, or create exploitable conditions in contract control flow.
\end{itemize}

\paragraph{\textbf{Exploitation Scenario:}}
\begin{itemize}
    \item An attacker could manipulate gas costs or fallback logic to make the call fail.
    \item The contract would continue under the false assumption that the Ether was transferred or the delegate executed safely.
    \item This allows denial of service or bypass of logical checks.
\end{itemize}

\paragraph{\textbf{Mitigation Recommendations:}}
\begin{itemize}
    \item Always verify the return value of low-level calls:
    \begin{lstlisting}[language=Solidity]
(bool success, ) = user.call{value: amount}("");
require(success, "Call failed");
    \end{lstlisting}

    \item Replace low-level calls with safer abstractions when possible, such as \texttt{transfer()}.
    \item Use explicit \texttt{revert()} or error handling when failures occur to maintain predictable behavior.
\end{itemize}

% ============================================================

\subsection{State and Storage Integrity Vulnerabilities}
\paragraph{This family concerns violations of storage consistency—when two pieces of code inadvertently share or corrupt the same storage context. 
Typical causes include unsafe use of \texttt{delegatecall}, inconsistent layout between proxy and implementation, or unchecked inline assembly writes. 
As indicated in Figure~\ref{fig:taxonomy-pdf}, storage-integrity faults form the boundary between logical correctness and authorization, often leading to full contract takeover.}

\subsubsection{DelegateCall Injection}\label{vuln:delegatecall}
\paragraph{\textbf{Definition:}}  
DelegateCall Injection arises when a smart contract uses \texttt{delegatecall} with an untrusted or user-controlled target. Since \texttt{delegatecall} runs the callee's code in the storage context of the calling contract, any injected logic can manipulate critical state variables or hijack control~\cite{pasqua2023enhancing, zhang2023svscanner}.

\begin{lstlisting}[caption={DelegateCall Injection Vulnerability}, label={lst:delegatecall}]
contract Proxy {
    address public implementation;

    function setImplementation(address _impl) public {
        implementation = _impl;  // Vulnerability: unprotected setter for delegatecall target
    }

    fallback() external payable {
        // Vulnerability: delegatecall without validation or access control
        (bool success, ) = implementation.delegatecall(msg.data);  
        // Vulnerability: raw msg.data forwarded without filtering
        // Vulnerability: no whitelist or function selector check
        require(success, "delegatecall failed");
    }
}
\end{lstlisting}

\paragraph{\textbf{Explanation of Vulnerability:}}

\begin{itemize}
    \item \textbf{Line 5 – Unprotected Target Setter }
    The \texttt{setImplementation()} function can be invoked by anyone, allowing replacement of the \texttt{implementation} contract with malicious logic.

    \item \textbf{Line 8 – Unsafe Use of delegatecall }
    The fallback function blindly uses \texttt{delegatecall} on user-controlled \texttt{msg.data} without function signature filtering, authentication, or validation.

    \item \textbf{Global – No Whitelist or Caller Control }  
    The proxy contract has no access control, condition checks, or contract validation mechanisms.
\end{itemize}

\paragraph{\textbf{Why This Is a Vulnerability:}} 
\begin{itemize}
  \item \textbf{Execution in caller’s storage context:}  
  The \texttt{delegatecall} instruction executes the callee’s code in the storage context of the caller. This means any state changes made by the callee directly affect the proxy’s storage, including critical variables such as ownership or balances.

  \item \textbf{Unrestricted target modification:}  
  Without access control on \texttt{setImplementation()}, any user can redirect the proxy to a malicious implementation, giving them full control over the proxy’s logic and state.

  \item \textbf{Arbitrary code execution:}  
  A malicious implementation can perform unauthorized actions such as overwriting privileged storage slots, draining Ether, or invoking \texttt{selfdestruct()}, effectively taking over or destroying the contract~\cite{pasqua2023enhancing, zhang2023svscanner}.

  \item \textbf{Lack of input validation and function filtering:}  
  The fallback function forwards raw \texttt{msg.data} without validating function selectors or ensuring that the target code is trusted. This enables attackers to craft arbitrary payloads for storage corruption or privilege escalation.

  \item \textbf{Immediate and irreversible impact:}  
  Because all changes occur within the proxy’s storage, exploitation results in direct and permanent corruption of the contract’s state—making recovery impossible once the malicious code executes.
\end{itemize}

\paragraph{\textbf{Exploitation Scenario:}}

\begin{itemize}
    \item An attacker deploys a malicious contract with logic to overwrite storage, drain funds, or selfdestruct.
    \item They invoke \texttt{setImplementation()} to point the proxy at the malicious contract.
    \item Using crafted \texttt{msg.data}, the attacker triggers arbitrary execution inside the proxy's context.
\end{itemize}

\paragraph{\textbf{Mitigation Recommendations:}}

\begin{itemize}
    \item Restrict \texttt{setImplementation()} using access control (e.g., \texttt{onlyOwner}).
    \item Validate new implementation contracts (e.g., check for minimum code size or trusted bytecode).
    \item Avoid forwarding raw \texttt{msg.data}; use explicit routing with known selectors.
    \item Follow hardened patterns like OpenZeppelin's \texttt{TransparentUpgradeableProxy}.
\end{itemize}

\subsubsection{Inline Assembly Misuse}\label{vuln:Inline Assembly Misuse}
\paragraph{\textbf{Definition:}}
\emph{Inline assembly misuse} refers to security faults introduced by using Yul/assembly to read or write memory/storage or to perform low-level calls without the type, bounds, and exception safety that Solidity normally enforces. Typical issues include unchecked \texttt{mload}/\texttt{calldataload}, arbitrary \texttt{sstore} to privileged slots, clobbering of storage layout, and ignoring call return values~\cite{chu2023survey, jiao2024survey, iuliano2024smart}.

\begin{lstlisting}[caption={Unsafe inline assembly Vulnerability}, label={lst:inline_asm_misuse}]
pragma solidity ^0.8.20;

contract AssemblyPitfalls {
    // Critical state
    address public owner;   // slot 0
    uint256 public balance; // slot 1

    constructor() { owner = msg.sender; }

    // 1) Unchecked read from calldata into arbitrary storage slot
    function setArbitrary(bytes calldata raw) external {
        assembly {
            // Vulnerability: no length check; reads past end if raw < 32 bytes
            let slot := calldataload(raw.offset)
            let val  := calldataload(add(raw.offset, 32))
            // Vulnerability: attacker-controlled sstore -> clobbers any slot (e.g., slot 0 owner)
            sstore(slot, val)
        }
    }

    // 2) Low-level call that ignores success and returndata
    function unsafeCall(address target, bytes calldata data) external {
        assembly {
            let ok := call(gas(), target, 0, data.offset, data.length, 0, 0)
            // Vulnerability: 'ok' ignored; execution continues even if target reverted
        }
    }

    // 3) Miscomputed slot via keccak; off-by-one corrupts adjacent mapping
    mapping(bytes32 => uint256) public limits; // layout uses keccak(key . slot)
    function setLimit(bytes32 key, uint256 v) external {
        assembly {
            // slot of 'limits' is 2, but we mistakenly use 3 -> writes foreign state
            mstore(0x00, key)
            mstore(0x20, 3)              // Vulnerability: wrong parent slot (should be 2)
            let slot := keccak256(0x00, 0x40)
            sstore(slot, v)
        }
    }
}
\end{lstlisting}

\paragraph{\textbf{Explanation of the vulnerability:}} 
Listing~\ref{lst:inline_asm_misuse} presents three misuse cases of inline assembly that compromise storage integrity and reliability.
\begin{itemize}
    \item \textbf{Lines 11–17 - } The function \texttt{setArbitrary()} performs unchecked \texttt{calldataload} operations on attacker-controlled input and directly executes \texttt{sstore(slot,val)}. This allows arbitrary writes to any storage slot, including critical variables such as \texttt{owner}.  
    \item \textbf{Lines 22–26 - } The function \texttt{unsafeCall()} executes a low-level \texttt{call} but ignores its success flag, enabling silent failures and masking reverted external executions.  
    \item \textbf{Lines 31–39 -} The function \texttt{setLimit()} miscomputes a mapping slot by writing to the wrong parent index (\texttt{3} instead of \texttt{2}), causing state overlap and corruption of unrelated variables.
\end{itemize}
Together, these errors show how small oversights in assembly—unchecked memory reads, ignored return values, and misaligned storage math—can escalate into full contract compromise.

\paragraph{\textbf{Why This is a Vulnerability:}}
\begin{compactitem}
  \item Assembly bypasses Solidity’s type checks, bounds checks, and storage layout guarantees; small mistakes yield arbitrary write primitives.
  \item Ignoring return values from low-level calls reintroduces failure-oblivious control flow.
  \item Miscomputed slots or offsets corrupt unrelated variables or mappings, breaking invariants and enabling privilege escalation~\cite{jiao2024survey, iuliano2024smart}.
\end{compactitem}

\paragraph{\textbf{Exploitation Scenario:}}
\begin{compactitem}
  \item An attacker calls \texttt{setArbitrary} with \texttt{slot=0} and \texttt{val=attacker}, resetting \texttt{owner}. They then take over administrative functions or drain funds. 
  \item Separately, \texttt{unsafeCall} can be used to mask downstream failures and leave accounting in an inconsistent state.
\end{compactitem}

\paragraph{\textbf{Mitigation Recommendations:}}
\begin{compactitem}
  \item Avoid inline assembly unless necessary; prefer safe Solidity primitives and libraries.
  \item If assembly is required, validate lengths/offsets before any \texttt{mload}/\texttt{calldataload}; encapsulate slot math in audited helpers and add invariants/comments.
  \item Always check and propagate the return value of \texttt{call}/\texttt{delegatecall}; bubble up revert reasons where possible.
  \item Lock down critical slots (e.g., use ERC-1967 fixed slots for proxy admin/implementation) and add runtime asserts that guard ownership variables.
  \item Use static analyzers and bytecode-level CFG/storage tools to catch slot math mistakes and unchecked calls~\cite{pasqua2023enhancing, ferreira2020smartbugs}.
\end{compactitem}

\subsubsection{Storage Overlap / Collision Attack}\label{vuln:Storage Overlap/Collision Attack}
\paragraph{\textbf{Definition:}}
A \emph{storage overlap/collision} occurs when contracts that share a storage context use inconsistent layouts, causing one contract to read/write slots assumed by another. In upgradeable patterns this can overwrite admin/owner pointers or corrupt application state~\cite{jiao2024survey, iuliano2024smart}.

\begin{lstlisting}[caption={Proxy storage collision Vulnerability}, label={lst:storage_collision}]
pragma solidity ^0.8.20;

// Minimal proxy: stores admin at slot 0, implementation at slot 1
contract MinimalProxy {
    address public admin;          // slot 0
    address public implementation; // slot 1

    constructor(address impl) { admin = msg.sender; implementation = impl; }

    fallback() external payable {
        address impl = implementation;
        assembly {
            calldatacopy(0, 0, calldatasize())
            let ok := delegatecall(gas(), impl, 0, calldatasize(), 0, 0)
            returndatacopy(0, 0, returndatasize())
            switch ok
            case 0 { revert(0, returndatasize()) }
            default { return(0, returndatasize()) }
        }
    }
}

// First implementation uses slot 0 for 'owner' -> collides with proxy.admin
contract ImplV1 {
    // !!! Collision: this variable sits at slot 0 in the shared storage
    address public owner; // slot 0 (overlaps MinimalProxy.admin)

    function init() external {
        owner = msg.sender; // Vulnerability: overwrites proxy.admin (slot collision)
    }

    function doAdminThing() external {
        require(msg.sender == owner, "not owner");
        // privileged action...
    }
}

// Safer layout: reserve slots or use ERC-1967 fixed slots + storage gaps
contract ImplV2 {
    // storage gap to avoid colliding with proxy slots (example pattern)
    uint256[50] private __gap;
    address public owner; // now at a higher, non-conflicting slot

    function init(address o) external { owner = o; }
}
\end{lstlisting}

\paragraph{\textbf{Explanation of Vulnerability:}}
In Listing~\ref{lst:storage_collision}, the vulnerability arises due to overlapping storage slots between the proxy and its implementation contract:
\begin{itemize}
  \item \textbf{Line 5 - Proxy Admin Declaration - } The \texttt{MinimalProxy} contract stores the \texttt{admin} variable at slot~0.
  \item \textbf{Line 26 - Implementation Variable Collision - } The \texttt{ImplV1} contract declares \texttt{owner} as its first state variable, which also occupies slot~0, colliding with \texttt{proxy.admin}.
  \item \textbf{Line 29 - Overwriting Critical Storage - } In \texttt{ImplV1.init()}, the statement \texttt{owner = msg.sender;} (see inline vulnerability comment) writes directly to slot~0 in the proxy’s storage, overwriting the admin’s address with the attacker’s.
  \item Because \texttt{delegatecall} on \textbf{Line 14} executes \texttt{ImplV1} in the proxy’s storage context, the overwrite occurs transparently, granting the attacker full administrative privileges.
\end{itemize}

\paragraph{\textbf{Why This is a Vulnerability:}}
\begin{compactitem}
  \item Shared storage with inconsistent layouts creates unintended aliasing between privileged pointers and application variables.
  \item Upgrades that reorder, insert, or change types of state variables shift slot indices, breaking invariants across versions.
  \item The effect is silent at compile time; the fault manifests only at runtime through \texttt{delegatecall}, often after deployment~\cite{olivieri2024software, iuliano2024smart}.
\end{compactitem}

\paragraph{\textbf{Exploitation Scenario:}}
An attacker triggers \texttt{ImplV1.init()} via the proxy (or any function that writes to slot~0), setting \texttt{owner=attacker} and thereby overwriting \texttt{proxy.admin}. With admin control, they upgrade to a malicious implementation and drain funds or brick the system.

\paragraph{\textbf{Mitigation Recommendations:}}
\begin{compactitem}
  \item Adopt ERC-1967/Transparent/UUPS patterns that store admin/implementation in fixed, hashed slots (e.g., \texttt{bytes32(uint256(keccak256("eip1967.proxy.admin"))-1)}).
  \item Preserve layout across upgrades: never reorder/remove variables; only append; use \emph{storage gaps} (\texttt{uint256[N] \_\_gap}) to future-proof layouts~\cite{jiao2024survey}.
  \item Add upgrade guards: access-controlled \texttt{onlyProxy}/\texttt{onlyAdmin}, code-hash/size checks on new implementations, and initialization modifiers that can be called only once.
  \item Employ static analysis and differential bytecode tools to compare layouts pre/post-upgrade and detect potential slot collisions~\cite{pasqua2023enhancing, ferreira2020smartbugs}.
  \item Consider formal specs for critical slots (admin/implementation/pauser) and assert them in on-chain getters or off-chain monitors to catch drift early~\cite{olivieri2024software}.
\end{compactitem}

\subsection{Arithmetic and Type Safety Vulnerabilities}
\paragraph{This family groups numerical and type-safety faults—overflows, underflows, and division-by-zero—that break arithmetic invariants. 
Before Solidity 0.8.0, arithmetic operations wrapped silently, enabling balance or time-lock bypasses. 
These issues underpin financial logic correctness and remain detectable patterns in legacy codebases.}

\subsubsection{Integer Overflow}\label{vuln:integer-overflow}
\paragraph{\textbf{Definition:}}  
An \emph{integer overflow} vulnerability occurs when an arithmetic operation on an unsigned or signed integer exceeds its maximum storable value. In Solidity versions prior to 0.8.0, this causes the value to wrap around to zero (modular arithmetic), potentially breaking contract logic or allowing unauthorized fund access~\cite{atzei2017survey, ferreira2020smartbugs, ghaleb2020effective, he2023detection, chu2023survey}.

\begin{lstlisting}[caption={integer overflow Vulnerability}, label={lst:int_overflow}]
pragma solidity >=0.4.21 <0.6.0;

contract DocumentSigner {
    mapping(address => uint) public lockTime_intou13;

    function increaseLockTime_intou13(uint _secondsToIncrease) public {
        // Vulnerability: addition without overflow check
        lockTime_intou13[msg.sender] += _secondsToIncrease; // overflow possible
    }

    function withdraw_intou13() public {
        require(now > lockTime_intou13[msg.sender]);
        uint transferValue_intou13 = 10;
        msg.sender.transfer(transferValue_intou13);
    }

    function bug_intou36(uint8 p_intou36) public {
        uint8 vundflw1 = 0;
        vundflw1 = vundflw1 + p_intou36; // overflow example
    }
}
\end{lstlisting}

\paragraph{\textbf{Explanation of Vulnerability:}}  
In Listing~\ref{lst:int_overflow} the root faults are concrete, local arithmetic writes that can be driven by attacker-controlled inputs and that silently wrap in pre-0.8.0 Solidity.  We point to the specific vulnerable lines and show inline vulnerability comments in the code (see listing) so readers and static tools can quickly locate the problem:
\begin{itemize}
  \item \textbf{Line 8 - Unchecked addition into storage (overflow primitive) }  
    \texttt{lockTime\_intou13[msg.sender] += \_secondsToIncrease;} is performed without any bounds check. If the stored value is near \texttt{type(uint).max} a sufficiently large \texttt{\_secondsToIncrease} will wrap the value modulo $2^{256}$, producing a small timestamp and effectively resetting the lock. (See the inline comment in the listing: \texttt{// Vulnerability: addition without overflow check}.)
  \item \textbf{Line 12 - Time comparison uses possibly-wrapped value }  
    \texttt{require(now > lockTime\_intou13[msg.sender]);} compares the current time against a storage slot that can be attacker-manipulated by the overflow at Line 8. After overflow the predicate can be made true immediately, bypassing the intended time-lock. (Also note: \texttt{now} is an alias for \texttt{block.timestamp} in older Solidity; the logical point is that a wrapped timestamp breaks the time check.)
  \item \textbf{Line 14 - Transfer after weakened check }  
    \texttt{msg.sender.transfer(transferValue\_intou13);} sends funds conditioned on the (now bypassed) time check. Because the prior arithmetic can force the require to pass, this transfer executes prematurely.
  \item \textbf{Line 19 - Small-width integer overflow example (local/stack) }  
    \texttt{vundflw1 = vundflw1 + p\_intou36;} demonstrates the same class of bug for an 8-bit variable: additions that exceed \texttt{uint8} wrap to small values, which in real contracts can translate into corrupted counters, allowance arithmetic, or token balances. (See inline comment: \texttt{// overflow example}.)
\end{itemize}

\paragraph{\textbf{Why This is a Vulnerability:}}  
\begin{itemize}
  \item Overflow enables breaking of logical invariants (e.g., time locks or balances) by manipulating arithmetic results.  
  \item Because Solidity versions prior to 0.8.0 perform unchecked arithmetic by default, overflowed values silently wrap around instead of reverting.  
  \item Attackers can exploit this to bypass authorization checks, manipulate storage variables, or unlock restricted resources~\cite{ferreira2020smartbugs, chu2023survey}.
\end{itemize}

\paragraph{\textbf{Exploitation Scenario:}}  
An attacker calls \texttt{increaseLockTime\_intou13()} with a large value such that \texttt{lockTime\_intou13[msg.sender]} overflows to a small number.  
The condition \texttt{now > lockTime\_intou13[msg.sender]} in \texttt{withdraw\_intou13()} then immediately evaluates to true, allowing premature withdrawal of locked Ether.  
This type of exploit has historically affected multiple time-lock and token contracts before the release of Solidity 0.8.0.

\paragraph{\textbf{Mitigation Recommendations:}}  
\begin{itemize}
  \item \textbf{Upgrade Solidity compiler:} Use Solidity \texttt{$\geq$0.8.0}, which automatically reverts on overflow and underflow.  
  \item \textbf{Use SafeMath library:} For older compilers, import \texttt{SafeMath} from OpenZeppelin to perform checked arithmetic:
\begin{lstlisting}[language=Solidity]
using SafeMath for uint256;
lockTime[msg.sender] = lockTime[msg.sender].add(_secondsToIncrease);
\end{lstlisting}
  \item \textbf{Validate inputs:} Restrict \texttt{\_secondsToIncrease} to reasonable bounds via \texttt{require()} statements.  
  \item \textbf{Perform fuzz testing:} Use automated tools to test arithmetic edge cases and verify no overflow occurs~\cite{ghaleb2020effective, he2023detection}.
\end{itemize}

\subsubsection{Integer Underflow}\label{vuln:integer-underflow}
\paragraph{\textbf{Definition:}}  
An \emph{integer underflow} occurs when a subtraction operation results in a value below zero in unsigned integer arithmetic.  
Before Solidity 0.8.0, this caused the value to wrap around to the maximum value representable by the type (e.g., $2^{256}-1$ for \texttt{uint256}), leading to unintended or exploitable outcomes~\cite{atzei2017survey, ferreira2020smartbugs, chu2023survey}.

\begin{lstlisting}[caption={integer underflow Vulnerability}, label={lst:int_underflow}]
pragma solidity >=0.4.21 <0.6.0;

contract DocumentSigner {
    function bug_intou39() public {
        uint8 vundflw = 0;
        vundflw = vundflw - 10;  // underflow bug
    }

    function bug_intou31() public {
        uint8 vundflw = 0;
        vundflw = vundflw - 10;  // underflow bug
    }

    function bug_intou35() public {
        uint8 vundflw = 0;
        vundflw = vundflw - 10;  // underflow bug
    }

    function bug_intou27() public {
        uint8 vundflw = 0;
        vundflw = vundflw - 10;  // underflow bug
    }
}
\end{lstlisting}

\paragraph{\textbf{Explanation of Vulnerability:}}  
In Listing~\ref{lst:int_underflow}, each function demonstrates an unchecked subtraction operation on unsigned integers, leading to an \emph{underflow} when the left operand is smaller than the right operand.  
Below we reference the precise vulnerable lines and explain the fault mechanism for each instance:

\begin{itemize}
  \item \textbf{Line 5 - }  
  The variable \texttt{vundflw} is initialized to \texttt{0} and then decremented by \texttt{10}.  
  The statement \texttt{vundflw = vundflw - 10;} (see inline comment \texttt{// underflow bug}) causes the value to wrap around from \texttt{0} to \texttt{246} (\texttt{2\textasciicircum8 - 10}) because unsigned integers cannot represent negative values.
  \item \textbf{Line 10 - }  
  The same operation is repeated: \texttt{vundflw = vundflw - 10;} subtracts from zero without any validation, again producing a large wrapped value rather than reverting.
  \item \textbf{Line 15 - }  
  The local variable underflows in the same way, representing a repeated unsafe arithmetic pattern typical of pre-0.8.0 Solidity codebases.
  \item \textbf{Line 20 - }  
  The final function mirrors the same issue; the subtraction is executed without bounds checking, creating a silent wrap-around that could, in real-world code, inflate balances or counters to unrealistic values.
\end{itemize} 
In pre-0.8.0 Solidity, such arithmetic does not revert, producing a large positive value rather than throwing an error.  
In practical contracts, similar operations on balances or counters can cause major logical inconsistencies or unauthorized token creation~\cite{ferreira2020smartbugs, ghaleb2020effective, chu2023survey}.

\paragraph{\textbf{Why This is a Vulnerability:}}  
\begin{itemize}
  \item Underflow can turn small values (like 0) into extremely large numbers, granting attackers unintended privileges or access to funds.  
  \item The silent wrap-around behavior enables bypassing of conditional checks and incorrect balance or time computations.  
  \item If exploited in payment or staking contracts, attackers may withdraw excessive funds or trigger logic that depends on low numeric thresholds.
\end{itemize}

\paragraph{\textbf{Exploitation Scenario:}}  
Suppose a balance tracking contract uses unsigned integers and allows subtraction without bounds checking:
\begin{lstlisting}[language=Solidity]
balances[msg.sender] -= withdrawalAmount;
\end{lstlisting}
If \texttt{balances[msg.sender]} is smaller than \texttt{withdrawalAmount}, the result underflows to a massive number, granting the attacker a huge balance and allowing them to drain the contract.  
In \texttt{DocumentSigner}, calling \texttt{bug\_intou39()} or \texttt{bug\_intou31()} demonstrates the same arithmetic wrap-around effect.

\paragraph{\textbf{Mitigation Recommendations:}}  
\begin{itemize}
  \item \textbf{Use Solidity 0.8.0 or higher:} Modern versions automatically revert on arithmetic underflow/overflow.  
  \item \textbf{Apply SafeMath subtraction:}
\begin{lstlisting}[language=Solidity]
using SafeMath for uint256;
balance = balance.sub(amount);
\end{lstlisting}
  \item \textbf{Enforce preconditions:} Always verify that operands maintain expected bounds before subtraction:
\begin{lstlisting}[language=Solidity]
require(balance >= amount, "Insufficient balance");
\end{lstlisting}
  \item \textbf{Static analysis and fuzz testing:} Integrate automated analyzers such as Mythril, Slither, or SmartBugs to detect integer underflow patterns early in the development cycle~\cite{ferreira2020smartbugs, he2023detection}.
\end{itemize}

\subsubsection{Division by Zero}\label{vuln:division-by-zero}
\paragraph{\textbf{Definition:}}  
A \emph{division-by-zero} vulnerability occurs when a smart contract performs an arithmetic division or modulo operation where the divisor is zero or can become zero due to user input or state manipulation.  
In Solidity versions prior to 0.8.0, such an operation causes a runtime exception and reverts the transaction, potentially halting dependent logic or enabling denial-of-service conditions.  
Although modern compilers revert safely, unvalidated user inputs or indirect zero divisors can still lead to broken logic, gas wastage, or DoS conditions~\cite{atzei2017survey, ferreira2020smartbugs, ghaleb2020effective, he2023detection, chu2023survey}.

\begin{lstlisting}[caption={Division-by-zero vulnerability}, label={lst:div_zero}]
pragma solidity ^0.5.0;

contract RewardDistributor {
    mapping(address => uint256) public rewards;
    uint256 public totalParticipants;

    function addParticipant(address participant, uint256 reward) public {
        rewards[participant] = reward;
        totalParticipants += 1;
    }

    function averageReward() public view returns (uint256) {
        // Vulnerability: possible division by zero
        // If no participants have been added, totalParticipants == 0
        return (address(this).balance) / totalParticipants;
    }

    function distribute() public {
        uint256 avg = averageReward(); // potential revert if divisor == 0
        // Distribution logic continues here...
    }
}
\end{lstlisting}

\paragraph{\textbf{Explanation of Vulnerability:}}  
In Listing~\ref{lst:div_zero} the bug is a single unchecked divisor used in a division operation — this is called a division-by-zero and will revert at runtime when the divisor equals zero. Key pointers:
\begin{itemize}
  \item \textbf{Line 15 - } \texttt{return (address(this).balance) / totalParticipants;} 
    performs division by \texttt{totalParticipants} without guaranteeing \texttt{> 0}.
  \item \textbf{Line 19 - } \texttt{uint256 avg = averageReward();}
     invoking the vulnerable line causes the call to revert and abort the transaction when \texttt{totalParticipants == 0}, allowing DoS of distribution logic.
  \item \textbf{Lines 7-10 - }
  \texttt{addParticipant(...)} increments \texttt{totalParticipants}, but there is no precondition ensuring it has been called before \texttt{averageReward()}, so the divisor can legitimately be zero.
\end{itemize}  
This halts execution and reverts the entire transaction, affecting all dependent logic and potentially freezing automated distribution systems.  
Static analyzers categorize this flaw under \emph{Arithmetic and Type Safety} vulnerabilities~\cite{ferreira2020smartbugs, he2023detection}.

\paragraph{\textbf{Why This is a Vulnerability:}}  
\begin{itemize}
  \item \textbf{Runtime exception:} Division or modulo by zero throws a runtime error that reverts state changes and consumes gas unnecessarily.  
  \item \textbf{Denial of Service (DoS):} Attackers or users can deliberately trigger zero divisors to halt execution and disrupt business logic or time-based reward systems.  
  \item \textbf{Inconsistent invariants:} Dependent calculations relying on unchecked divisors (e.g., averages, ratios, or percentages) can produce undefined or reverted results, undermining contract reliability~\cite{chu2023survey}.
\end{itemize}

\paragraph{\textbf{Exploitation Scenario:}}  
Consider a reward or staking contract where an administrative function calls \texttt{averageReward()} periodically.  
If an attacker invokes \texttt{distribute()} before any participants are registered (i.e., \texttt{totalParticipants == 0}), the division operation reverts and halts the transaction.  
Repeated exploitation can render the contract temporarily unusable, blocking reward distribution or freezing dependent automated jobs.  
In older smart contracts, such behavior could even cause gas exhaustion loops in dependent contracts that call the vulnerable function.

\paragraph{\textbf{Mitigation Recommendations:}}  
\begin{itemize}
  \item \textbf{Validate divisor inputs:} Always check that divisors are non-zero before performing division or modulo operations:
\begin{lstlisting}[language=Solidity]
require(totalParticipants > 0, "Division by zero");
uint256 average = totalRewards / totalParticipants;
\end{lstlisting}
  \item \textbf{Apply defensive programming:} Use conditional returns to handle zero cases gracefully instead of letting execution revert:
\begin{lstlisting}[language=Solidity]
if (totalParticipants == 0) return 0;
\end{lstlisting}
  \item \textbf{Use Solidity 0.8.0 or higher:} Modern compilers automatically revert on division-by-zero but validation should still be explicit for logical clarity and to avoid revert-based DoS chains.  
  \item \textbf{Perform thorough unit and fuzz testing:} Test arithmetic functions with boundary values (e.g., divisor = 0) to ensure contracts handle zero divisions gracefully~\cite{ghaleb2020effective, he2023detection}.
  \item \textbf{Static and dynamic analysis:} Include this check in automated pipelines (Mythril, Slither, SmartBugs) as part of the Arithmetic Safety rule set.
\end{itemize}

% ============================================================

\subsection{Blockchain-Environment Dependency Vulnerabilities}
\paragraph{Contracts that depend directly on mutable blockchain parameters—such as timestamps, block numbers, or difficulty—inherit the uncertainty of those values.This family captures vulnerabilities where environment-controlled data influence critical decisions like randomness or deadlines, enabling subtle but systematic bias by miners or validators.}

\subsubsection{Timestamp Dependency}\label{vuln:timestamp}
\paragraph{\textbf{Definition:}}
Timestamp dependence occurs when a smart contract uses \texttt{block.timestamp} in critical decision-making, such as reward distribution or game deadlines. Since miners can slightly manipulate timestamps within a permissible range, attackers may influence contract logic in their favor~\cite{qian2023demystifying}.

\begin{lstlisting}[caption={Timestamp Dependency Vulnerability}, label={lst:timestamp}]
contract Lottery {
    address public winner;
    uint public closeTime;

    constructor() {
        closeTime = block.timestamp + 1 days;
    }

    function drawWinner() public {
        require(block.timestamp >= closeTime, "Too early");
        winner = msg.sender;  // Vulnerability: anyone can call at the right time
    }
}
\end{lstlisting}

\paragraph{\textbf{Explanation of Vulnerability:}}
\begin{itemize}
    \item \textbf{Line 10 - } The function \texttt{drawWinner()} depends on \texttt{block.timestamp} to determine whether it is time to finalize the lottery. However, miners can influence the timestamp by several seconds, enabling unfair advantage.
    \item \textbf{Line 11 - } There is no randomness or user verification—any account that calls the function after the threshold becomes the winner.
\end{itemize}

\paragraph{\textbf{Why This Is a Vulnerability:}}
\begin{itemize}
    \item Miners can include this transaction in a block with a biased \texttt{timestamp}, satisfying the \texttt{require()} condition earlier or later than intended.
    \item In timing-critical applications (e.g., lotteries, auctions), this subtle control creates an unfair and manipulable environment.
\end{itemize}

\paragraph{\textbf{Exploitation Scenario:}}
\begin{itemize}
    \item A miner, or an attacker working with one, monitors the block's timestamp range and submits a transaction when the contract will permit it.
    \item By controlling inclusion time, the attacker ensures they are the first to invoke \texttt{drawWinner()} and claim rewards.
\end{itemize}

\paragraph{\textbf{Mitigation Recommendations:}}
\begin{itemize}
    \item Avoid using \texttt{block.timestamp} for critical control logic, especially where fairness or randomness is involved.
    \item For randomness, use verifiable off-chain sources like Chainlink VRF.
    \item For deadlines, consider tolerating timestamp manipulation or using block numbers with defined intervals.
\end{itemize}

\subsubsection{Block Information Dependency}\label{vuln:Block Information Dependency}
\paragraph{\textbf{Definition:}}  
This vulnerability arises when developers attempt to generate randomness using predictable blockchain variables such as \texttt{block.timestamp}, \texttt{block.difficulty}, or \texttt{blockhash}. These values can be influenced or anticipated by miners and are therefore insecure sources for randomness in smart contracts~\cite{qian2023demystifying, sun2023assbert}.

\begin{lstlisting}[caption={Block Information Dependency}, label={lst:randomness}]
contract RandomnessBlockInfo {
    function useTimestamp() public view returns (uint) {
        // Vulnerability: block.timestamp can be manipulated by miners
        return uint(keccak256(abi.encodePacked(block.timestamp)));
    }

    function useMultipleBlockVars() public view returns (uint) {
        // Vulnerability: using predictable values like block.difficulty and block.coinbase
        return uint(keccak256(abi.encodePacked(block.difficulty, block.coinbase)));
    }

    function useBlockhash(uint index) public view returns (bytes32) {
        // Vulnerability: blockhash is deterministic and not suitable for randomness
        return blockhash(block.number - index);
    }
}
\end{lstlisting}

\paragraph{\textbf{Explanation of Vulnerability:}} 
 
This contract demonstrates three distinct insecure practices:

\begin{itemize}
    \item \textbf{Line 4 – Use of \texttt{block.timestamp} }Miners can manipulate the block timestamp slightly to influence outcomes, especially in games and lotteries.
    \item \textbf{Line 9 – Use of \texttt{block.difficulty} and \texttt{block.coinbase} }These values are also predictable and do not offer entropy suitable for randomness.
    \item \textbf{Line 14 – Use of \texttt{blockhash} } The blockhash of past blocks is publicly available and deterministic. It does not provide true unpredictability.
\end{itemize}

\paragraph{\textbf{Why This is a Vulnerability:}}
Blockchain state variables such as \texttt{block.timestamp}, \texttt{block.difficulty}, and \texttt{blockhash} are accessible to both the contract and the miner but are neither cryptographically random nor immutable during mining. This predictability permits strategic manipulation by miners or attackers to bias outcomes that rely on these values. In randomness-sensitive applications—like lotteries, auctions, and entropy-based minting—this breaks the fairness guarantees expected by participants and may cause economic loss~\cite{qian2023demystifying, cai2023combine}.

\paragraph{\textbf{Exploitation Scenario:}}
\begin{itemize}
    \item A smart contract uses \texttt{block.timestamp} or \texttt{blockhash} to select a reward recipient.
    \item A miner observes the contract logic and chooses to include or exclude certain transactions to bias the outcome.
    \item The miner manipulates the timestamp or delays a block to maximize their chance of winning.
\end{itemize}

\paragraph{\textbf{Mitigation Recommendations:}}
\begin{itemize}
    \item Avoid using blockchain properties for randomness in production logic.
    \item Use secure external sources such as Chainlink VRF for verifiable randomness.
    \item Document any pseudo-random logic clearly if it is non-critical or only used for display/UI purposes.
\end{itemize}

\subsubsection{Bad Randomness}\label{vuln:Bad Randomness}
\paragraph{\textbf{Definition:}}
Bad randomness refers to the use of insecure or manipulable values—typically derived from blockchain state variables—for generating random numbers. Since these values are either publicly known or subject to proposer/miner influence, they cannot guarantee unpredictability or fairness in randomized smart contract logic~\cite{qian2023demystifying, sun2023assbert}.

\begin{lstlisting}[caption={Bad randomness Vulnerability}, label={lst:bad_randomness}]
pragma solidity ^0.4.0;
contract Lottery {
    event GetBet(uint betAmount, uint blockNumber, bool won);
    struct Bet {
        uint betAmount;
        uint blockNumber;
        bool won;
    }
    address private organizer;
    Bet[] private bets;
    function Lottery() { organizer = msg.sender; }
    function() { throw; } // legacy revert
    // Make a bet
    function makeBet() {
        // Won if block number is even (public, miner-influenced; not random)
        // <yes> <report> BAD_RANDOMNESS
        bool won = (block.number % 2) == 0;            // (Line 17) Vulnerability: predictable "randomness"
        // Record the bet with an event
        // <yes> <report> BAD_RANDOMNESS
        bets.push(Bet(msg.value, block.number, won));  // (Line 20) Uses same manipulable source in outcome log
        if (won) {
            if (!msg.sender.send(msg.value)) {
                throw;
            }
        }
    }
    function getBets() {
        if (msg.sender != organizer) { throw; }
        for (uint i = 0; i < bets.length; i++) {
            GetBet(bets[i].betAmount, bets[i].blockNumber, bets[i].won);
        }
    }
    function destroy() {
        if (msg.sender != organizer) { throw; }
        suicide(organizer);
    }
}
\end{lstlisting}

\paragraph{\textbf{Explanation of Vulnerability:}}
\begin{itemize}
  \item \textbf{Line 17 – Predictable entropy } The outcome is derived from \texttt{block.number \% 2}. Block numbers are public and known at inclusion time; proposers/miners can influence which transactions land in even vs.\ odd blocks.
  \item \textbf{Line 20 – Outcome tied to same block } The contract records \texttt{block.number} alongside the bet, cementing a direct link between the manipulable variable and the payout decision in the same transaction.
  \item \textbf{Single-transaction finality } No commit–reveal or delay separates the user’s input from the entropy sample, enabling atomic manipulation or strategic reordering.
\end{itemize}

\paragraph{\textbf{Why This is a Vulnerability:}}
\begin{itemize}
  \item \textbf{Public and miner-influenced inputs:} \texttt{block.number}, \texttt{block.timestamp}, \texttt{blockhash}(\(n\)) (for small \(n\)), and similar fields are observable and, within limits, biasable by block proposers/validators.
  \item \textbf{Predictability breaks fairness:} Adversaries can forecast outcomes before confirmation and choose when to submit, cancel, or reorder transactions to gain a systematic edge.
  \item \textbf{Atomic exploitation:} Without a delayed reveal or external verifiable randomness, attackers can shape inclusion (e.g., target even blocks) and immediately capture favorable payouts.
\end{itemize}

\paragraph{\textbf{Exploitation Scenario:}}
\begin{itemize}
  \item \textbf{Gas-price frontrun to even blocks:} An attacker observes that even block numbers win. They resubmit their bet with a higher gas price when the mempool suggests the next block will be even, increasing the chance their bet executes in a winning block.
  \item \textbf{Proposer/miner bias:} A block proposer places their own bet transactions only into even-numbered blocks (or withholds/defers inclusion) to harvest payouts reliably.
  \item \textbf{MEV relay strategy:} An MEV bot sandwiches user bets to ensure its own transaction lands in a favorable parity block while pushing others into losing ones.
\end{itemize}

\paragraph{\textbf{Mitigation Recommendations:}}
\begin{itemize}
  \item \textbf{Use verifiable randomness (preferred):} Integrate a VRF oracle (e.g., on-chain VRF) to obtain an unpredictable, publicly verifiable random value.
  \item \textbf{Commit–reveal with delay:} Require users to commit a hashed secret first and reveal after \(k\) blocks. Combine user salt with future, \emph{not yet known} entropy to reduce proposer bias.
  \item \textbf{Separate phases across blocks:} Sample entropy from a block \emph{after} the bet is irrevocably recorded (e.g., use block \(n{+}k\)) to limit same-block manipulation; avoid direct use of \texttt{block.timestamp}/\texttt{block.number}.
  \item \textbf{Aggregate multi-party entropy:} Mix multiple independent sources (user commits + VRF + operator beacons) so that no single party can unilaterally bias the output.
  \item \textbf{Audit outcome linkage:} Ensure payout logic references only verifiable, unbiased randomness and that logs/events do not encode manipulable “random” fields as authoritative outcomes.
\end{itemize}

\subsection{Access Control and Authentication Vulnerabilities}
\paragraph{This family addresses failures in privilege enforcement—when ownership or authorization checks are missing, inconsistent, or incorrectly implemented. Such flaws directly expose administrative or monetary functions to unauthorized users. 
They represent one of the most common and severe root causes across DeFi incidents.}

\subsubsection{Access Control Misconfiguration}\label{vuln:accesscontrol}
\paragraph{\textbf{Definition:}}  
An \emph{access control misconfiguration} vulnerability occurs when a smart contract fails to enforce or improperly implements authorization checks on sensitive functions.  
Such misconfigurations allow unauthorized users to perform privileged actions—such as transferring Ether, modifying critical state variables, or bypassing restrictions—that should be reserved for the contract owner or designated roles~\cite{atzei2017survey, ferreira2020smartbugs, ghaleb2020effective, chu2023survey, he2023detection}.  
Common causes include missing \texttt{require()} statements, incorrect use of ownership patterns, or omission of balance resets and state updates after privileged operations.

\begin{lstlisting}[caption={Access control Vulnerability}, label={lst:access_control}]
pragma solidity ^0.4.24;
contract Wallet {
    address creator;
    mapping(address => uint256) balances;

    constructor() public {
        creator = msg.sender;
    }
    function deposit() public payable {
        assert(balances[msg.sender] + msg.value > balances[msg.sender]);
        balances[msg.sender] += msg.value;
    }
    function withdraw(uint256 amount) public {
        require(amount <= balances[msg.sender]);
        msg.sender.transfer(amount);
        balances[msg.sender] -= amount;
    }
    function refund() public {
        // <yes> <report> ACCESS_CONTROL
        // Vulnerability: no authorization or balance reset
        msg.sender.transfer(balances[msg.sender]);
    }
    function migrateTo(address to) public {
        // Only creator can migrate funds
        require(creator == msg.sender);
        to.transfer(this.balance);
    }
}
\end{lstlisting}

\paragraph{\textbf{Explanation of Vulnerability:}}
\begin{itemize}
  \item \textbf{Lines 18–22 — Unprotected refund path -} 
  The function \texttt{refund()} (Line~18) lacks any authorization or role check before transferring funds. 
  The inline marker (Line~19) flags this as \texttt{ACCESS\_CONTROL}, yet no \texttt{require(...)} is present to restrict callers.

  \item \textbf{Line 21 — Missing state update after transfer -} 
  \texttt{msg.sender.transfer(balances[msg.sender])} sends Ether but \emph{does not} reset \texttt{balances[msg.sender]} to zero. 
  This enables repeated calls to \texttt{refund()} to withdraw the same recorded balance multiple times.

  \item \textbf{Missing line (after Line 21) — Absent balance reset - } 
  A critical line such as \texttt{balances[msg.sender] = 0;} is missing immediately after the transfer, leaving the per-user credit intact and enabling unlimited refunds.

  \item \textbf{Policy inconsistency (Lines 23–26) vs.\ refund (Lines 18–22) - } 
  \texttt{migrateTo()} correctly enforces \texttt{require(creator == msg.sender)} (Line~25), but \texttt{refund()} applies no similar check for a fund-moving operation, creating an inconsistent access policy.
\end{itemize}

Consequently, any caller can repeatedly drain Ether via \texttt{refund()} due to missing authorization and missing post-transfer balance reset, leading to complete contract depletion.
 
Furthermore, although \texttt{migrateTo()} correctly restricts access to the contract creator, the absence of similar access validation in \texttt{refund()} creates a partial protection gap~\cite{ferreira2020smartbugs, ghaleb2020effective}.

\paragraph{\textbf{Why This is a Vulnerability:}}  
\begin{itemize}
  \item \textbf{Missing state update:} The user’s balance is not reset after transferring funds, allowing repeated withdrawals.  
  \item \textbf{Improper privilege enforcement:} Any account can call \texttt{refund()} without requiring ownership or authorization.  
  \item \textbf{Inconsistent access policy:} The contract enforces access control in one function (\texttt{migrateTo()}) but neglects it in another that performs critical fund transfers.  
  \item \textbf{Financial loss:} This leads to unlimited refunds and complete depletion of the contract’s Ether balance~\cite{chu2023survey, he2023detection}.
\end{itemize}

\paragraph{\textbf{Exploitation Scenario:}}  
An attacker deposits a small amount of Ether (e.g., 1 wei) using \texttt{deposit()}, which credits their balance.  
They then repeatedly invoke \texttt{refund()} to transfer their full recorded balance each time, without the contract ever resetting it to zero.  
Each invocation drains an equal amount from the contract’s total Ether holdings.  
Since the function lacks both state mutation and access restrictions, a single attacker can fully exhaust the wallet’s balance.  
This vulnerability was observed in multiple early Ethereum wallets and crowdsale contracts.

\paragraph{\textbf{Mitigation Recommendations:}}  
\begin{itemize}
  \item \textbf{Reset state after critical actions:} Always update user balances or access flags before or immediately after performing fund transfers:
\begin{lstlisting}[language=Solidity]
uint256 amount = balances[msg.sender];
balances[msg.sender] = 0;
msg.sender.transfer(amount);
\end{lstlisting}
  \item \textbf{Use ownership and role-based access control:}  
  Employ the OpenZeppelin \texttt{Ownable} or \texttt{AccessControl} modules to explicitly define administrative privileges:
\begin{lstlisting}[language=Solidity]
modifier onlyOwner() {
    require(msg.sender == owner, "Not authorized");
    _;
}
\end{lstlisting}
  \item \textbf{Follow the Checks–Effects–Interactions (CEI) pattern:}  
  Verify preconditions, update internal state, and only then interact with external addresses.  
  This prevents reentrancy and ensures correct state before external transfers.  
  \item \textbf{Comprehensive testing and auditing:}  
  Use automated analyzers (e.g., Slither, SmartBugs, Mythril) and manual audits to verify consistent access control across all critical functions~\cite{ferreira2020smartbugs, ghaleb2020effective, he2023detection}.  
  \item \textbf{Consistent design policy:}  
  Define a unified access policy for the contract, ensuring all fund-moving or state-changing functions follow the same authentication and update logic.
\end{itemize}

\textbf{Notes:}  
Access control misconfigurations remain one of the most prevalent and high-impact vulnerabilities in smart contracts.  
Even simple omissions—such as forgetting to reset balances or missing a \texttt{require(msg.sender == owner)} statement—can lead to full asset compromise.  
Modern frameworks and static analyzers include explicit detection heuristics for missing state updates and inconsistent access validation.

\subsubsection{tx.origin Misuse}\label{vuln:txorigin}
\paragraph{\textbf{Definition:}}
The misuse of \texttt{tx.origin} occurs when a contract relies on this global variable to authorize users. Unlike \texttt{msg.sender}, which refers to the immediate caller, \texttt{tx.origin} refers to the original external account that initiated the transaction. This makes it vulnerable to phishing-style attacks, where a malicious contract tricks a user into triggering sensitive logic in another contract~\cite{zhang2022cbgru, chu2023survey}.

\begin{lstlisting}[caption={tx.origin Vulnerability}, label={lst:txorigin}]
contract TxOriginExample {
    address public owner;
    constructor() {
        owner = msg.sender;
    }
    function transferOwnership(address newOwner) public {
        // Vulnerability: insecure use of tx.origin for authorization
        require(tx.origin == owner, "Not authorized");
        owner = newOwner;
    }
    function safeTransfer(address newOwner) public {
        // Warning: combining msg.sender and tx.origin still unsafe
        require(msg.sender == owner && tx.origin == owner, "Still unsafe pattern");
        owner = newOwner;
    }
}
\end{lstlisting}

\paragraph{\textbf{Explanation of Vulnerability:}}
\begin{itemize}
    \item \textbf{Line 8 - } The function \texttt{transferOwnership} uses \texttt{require(tx.origin == owner)} for access control. If a malicious contract indirectly invokes this function while the victim is the transaction origin, the check passes—despite the actual caller being malicious.
    \item \textbf{Line 13 - } Even though \texttt{safeTransfer} uses both \texttt{msg.sender} and \texttt{tx.origin}, the inclusion of \texttt{tx.origin} still makes it unsafe. This line triggers \texttt{heuristic\_tx\_origin\_check}.
\end{itemize}

\paragraph{\textbf{Why This Is a Vulnerability:}}
\begin{itemize}
    \item The use of \texttt{tx.origin} exposes the contract to phishing-like attacks. A malicious contract can be designed to execute the vulnerable function while the user unknowingly acts as the transaction origin.
    \item Since \texttt{tx.origin} cannot distinguish between trusted and untrusted intermediate calls, it should never be used in critical logic such as ownership validation.
\end{itemize}

\paragraph{\textbf{Exploitation Scenario:}}
\begin{itemize}
    \item An attacker deploys a contract that calls \texttt{transferOwnership}.
    \item The victim interacts with this attacker contract. Because the victim’s wallet is the transaction origin, the check on Line 9 passes.
    \item Ownership is silently transferred to the attacker without the victim's knowledge.
\end{itemize}

\paragraph{\textbf{Mitigation Recommendations:}}
\begin{itemize}
    \item Always use \texttt{msg.sender} for authorization checks to validate the direct caller.
    \item Never use \texttt{tx.origin} in ownership or sensitive access logic.
    \item Use established security libraries such as \texttt{Ownable} from OpenZeppelin to handle ownership safely.
\end{itemize}

\subsubsection{Unauthorized Selfdestruct (Suicidal Contract)}\label{vuln:suicide}
\paragraph{\textbf{Definition:}}
A suicidal contract vulnerability arises when a contract includes a publicly accessible or improperly restricted \texttt{selfdestruct} call, allowing unauthorized users to irreversibly terminate the contract and redirect any remaining Ether~\cite{ferreira2020smartbugs, colin2024integrated}.

\begin{lstlisting}[caption={Unprotected Selfdestruct Vulnerability}, label={lst:suicide}]
contract SuicideContract {
    address public owner;
    constructor() {
        owner = msg.sender;
    }
    function destroy() public {
        // Vulnerability: selfdestruct is called without checking ownership
        selfdestruct(payable(msg.sender));
    }
}
\end{lstlisting}

\paragraph{\textbf{Explanation of the Vulnerability:}}

\begin{itemize}
    \item \textbf{Line 8 – Selfdestruct without access contro - } 
    The \texttt{destroy()} function is publicly accessible and allows any caller to execute \texttt{selfdestruct}. It transfers the entire contract balance to the caller (\texttt{msg.sender}) without verifying if the caller is the owner. This violates fundamental access control expectations for critical operations.
\end{itemize}

\paragraph{\textbf{Why This Is a Vulnerability:}}
\begin{itemize}
    \item The contract can be terminated at any time by any address.
    \item Ether stored in the contract is transferred to the attacker.
    \item Contract functionality is destroyed permanently, disrupting dependent applications and users.
\end{itemize}

\paragraph{\textbf{Exploitation Scenario:}}
\begin{itemize}
    \item An attacker discovers the \texttt{destroy()} function and calls it.
    \item The contract is terminated, and any remaining funds are sent to the attacker.
    \item All future calls to the contract fail, since the bytecode is removed from the blockchain.
\end{itemize}

\paragraph{\textbf{Mitigation Recommendations:}}
\begin{itemize}
    \item Enforce strict access control before calling \texttt{selfdestruct}. For example:
\begin{lstlisting}[language=Solidity]
require(msg.sender == owner, "Not authorized");
\end{lstlisting}
    \item Use the \texttt{onlyOwner} modifier or similar role-based access control for destructive functions.
    \item Avoid using \texttt{selfdestruct} unless contract destruction is part of the design. Prefer disabling functionality via flags.
\end{itemize}

% ============================================================

\subsection{ABI and Input Validation Vulnerabilities}
\paragraph{At the interface between users and contract logic, improper validation of calldata or encoded parameters can shift argument boundaries or allow ambiguous interpretation. These vulnerabilities manifest as short-address bugs, unsafe \texttt{bytes} decoding, or hash collisions caused by packed encodings.}

\subsubsection{Short Address Attack}\label{vuln:shortaddress}
\paragraph{\textbf{Definition:}}
Short Address Attack exploits Ethereum's padding and alignment behavior for function arguments, particularly when address-type inputs are passed with fewer than 20 bytes, causing subsequent parameters to be misinterpreted~\cite{ghaleb2020effective}.
\begin{lstlisting}[caption={Short Address Vulnerability}, label={lst:shortaddress}]
pragma solidity ^0.4.24;
contract ShortAddressAttackExamples {
    mapping(address => uint256) public balances;
    // Classic short address vulnerability using raw address
    function transfer(address to, uint256 amount) public {
        balances[to] += amount;  // Vulnerability: No input length check for 'to' (short address attack)
        balances[msg.sender] -= amount;
    }
    // Uses bytes input assumed to contain an address
    function unsafeSetAddress(bytes memory raw) public {
        address extracted;
        assembly {
            extracted := mload(add(raw, 20))  // Vulnerability: Unsafe use of mload on unchecked bytes
        }
        balances[extracted] = 100;
    }
    // Uses an address input in a call without checking length
    function callReceiver(address receiver) public {
        receiver.call("");  // Vulnerability: Unsafe call to address without validating input integrity
    }
    // Safe version with length check
    function safe(bytes memory raw) public {
        require(raw.length == 20, "Invalid length");
        address extracted;
        assembly {
            extracted := mload(add(raw, 20))
        }
        balances[extracted] = 123;
    }
}
\end{lstlisting}

\paragraph{\textbf{Explanation of Vulnerabilities:}}
\begin{itemize}
    \item \textbf{Line 6 – Raw \texttt{address} parameter without length check } 
    The \texttt{transfer} function uses the \texttt{to} address parameter directly. When encoded calldata is too short, Solidity may misalign the address and \texttt{uint256} parameters, causing fund misdirection.

    \item \textbf{Line 13 – Unsafe assembly with unchecked \texttt{bytes} }
    The function assumes \texttt{raw} contains at least 20 bytes but does not check this before accessing memory directly via \texttt{mload}. This introduces undefined behavior and may extract incorrect addresses.

    \item \textbf{Line 19 – External call to unchecked address }
    The \texttt{receiver} parameter is used in a low-level call without confirming whether the address is properly constructed or safe.

    \item \textbf{Line 22 – Secure implementation with length validation }
    Serves as a best practice: explicitly checks that \texttt{raw.length == 20} to ensure integrity before extracting an address.
\end{itemize}

\paragraph{\textbf{Why This Is a Vulnerability:}}
\begin{itemize}
    \item Ethereum pads calldata on the right, and improper input length causes offsets to shift unexpectedly.
    \item Attackers can exploit this by crafting calldata that causes value misalignment or function argument corruption.
    \item Inline assembly makes the attack surface worse if bytes are not checked before memory reads.
\end{itemize}

\paragraph{\textbf{Exploitation Scenario:}}
\begin{itemize}
    \item An attacker sends a 19-byte address followed by carefully crafted values, causing the address and amount parameters in \texttt{transfer} to be interpreted incorrectly.
    \item This can result in tokens being sent to the wrong address or balance underflows that favor the attacker.
    \item If unchecked bytes are passed to \texttt{unsafeSetAddress}, it may silently write balance to unintended addresses.
\end{itemize}

\paragraph{\textbf{Mitigation Recommendations:}}
\begin{itemize}
    \item Always validate the length of \texttt{bytes} before extracting addresses from them:
\begin{lstlisting}[language=Solidity]
require(raw.length == 20, "Invalid address length");
\end{lstlisting}

    \item Avoid direct memory reads using inline assembly unless absolutely required.
    \item Upgrade to modern Solidity versions ($>=~0.5.0$) which enforce stricter input checking.
    \item Use well-tested libraries for encoding/decoding calldata or low-level operations.
\end{itemize}

% ============================================================

\subsubsection{Packed-ABI Hash Collision}\label{vuln:packed-abi-collision}
\paragraph{\textbf{Definition:}}
When hashing concatenated variable-length fields via \texttt{abi.encodePacked}, distinct tuples can serialize to identical byte streams, yielding the same hash and enabling intent confusion or signature/commit forgery~\cite{chu2023survey, vidal2024vulnerability}.

\begin{lstlisting}[caption={Packed-ABI Hash Collision Vulnerability},label={lst:packed-collision}]
pragma solidity ^0.8.0;
contract PackedABICollision {
    mapping(bytes32 => bool) public executed;
    // Vulnerable: variable-length fields packed without separators/types
    function authorize(string memory action, string memory param) external {
        // Collisions: ("ab","c") vs ("a","bc") -> same packed bytes
        // <yes> <report> PACKED_ABI_COLLISION
        bytes32 h = keccak256(abi.encodePacked(action, param)); // Vulnerability: ambiguous packing
        require(!executed[h], "already used");
        executed[h] = true;
        // ... perform action "authorized" by (action,param)
    }
    // Safer alternative:
    function authorizeTyped(string memory action, string memory param) external {
        // abi.encode is typed/length-prefixed -> no such collisions
        bytes32 h = keccak256(abi.encode(action, param));
        require(!executed[h], "already used");
        executed[h] = true;
    }
}
\end{lstlisting}

\paragraph{\textbf{Explanation of Vulnerabilities}}
\begin{itemize}
  \item \textbf{Line 8 — Ambiguous concatenation } \texttt{abi.encodePacked(action,param)} elides type and length prefixes; variable-length strings can produce identical concatenations, e.g., \((\texttt{"ab"},\texttt{"c"})\) and \((\texttt{"a"},\texttt{"bc"})\) serialize to the same bytes.
  \item \textbf{Line 8 — Colliding hash as authority key } The contract derives \texttt{h} from the ambiguous packed bytes and treats it as an authorization/uniqueness key, so different input tuples can map to the same \texttt{h}.
  \item \textbf{Lines 8–9 — Logic bound to \texttt{h} } Both the replay check and state update hinge solely on \texttt{h}; a collision lets an attacker pass the check with semantically different parameters or pre-burn a victim’s hash.
  \item \textbf{Scope:} The issue generalizes beyond strings to any mix of variable-length fields (e.g., \texttt{bytes}, \texttt{string}) and even some fixed-width concatenations when not delimited with domain separators.
\end{itemize}

\paragraph{\textbf{Why This is a Vulnerability:}}
\begin{itemize}
  \item \textbf{Input ambiguity undermines binding:} The hash no longer uniquely binds the intended tuple \((\texttt{action},\texttt{param})\); distinct inputs can authorize the same effect.
  \item \textbf{Second-preimage feasibility:} Finding an alternate tuple that packs to an existing byte stream is trivial (string splicing, crafted \texttt{bytes}), enabling commit forgery and signature misuse.
  \item \textbf{State/key collisions:} Using the colliding \texttt{h} as a mapping key allows preemption (burning) or impersonation of another tuple’s authorization.
  \item \textbf{Cross-domain confusion:} Reusing packed hashes across subsystems (orders, roles, claims) without domain separation enables unintended replay or privilege escalation.
\end{itemize}

\paragraph{\textbf{Exploitation Scenario:}}
\begin{itemize}
  \item \textbf{Parameter substitution:} A victim intends to authorize \((\texttt{"ab"},\texttt{"c"})\). The attacker submits \((\texttt{"a"},\texttt{"bc"})\), which produces the same \texttt{h} (Line~7), passing the uniqueness check (Line~8) and executing unintended logic.
  \item \textbf{Pre-burn (DoS):} The attacker first calls \texttt{authorize("a","bc")} to set \texttt{executed[h]=true} (Line~9), preventing the victim’s later, legitimate \texttt{authorize("ab","c")}.
  \item \textbf{Signature replay across intents:} If an off-chain signature covers \(\mathtt{keccak256(abi.encodePacked(action,param))}\), the attacker replaces the signed tuple with a colliding pair and reuses the signature to authorize a different action.
\end{itemize}

\paragraph{\textbf{Mitigation Recommendations:}}
\begin{itemize}
  \item \textbf{Use typed encoding:} Prefer \texttt{keccak256(abi.encode(...))} so type and length prefixes remove ambiguity (see \texttt{authorizeTyped}).
  \item \textbf{Add domain separation:} Prefix encodings with a constant tag and version (e.g., \texttt{keccak256(abi.encode("AUTH\_V1", action, param))}).
  \item \textbf{Avoid packing multiple dynamic fields:} If \texttt{abi.encodePacked} must be used, do not concatenate two or more dynamic types; alternatively, include explicit length delimiters for each dynamic field.
  \item \textbf{Ethereum Improvement Proposal (EIP)-712 for signatures:} Use typed structured data with domain separator and explicit field types when producing/verifying off-chain signatures.
  \item \textbf{Hash fields individually then combine:} For dynamic fields, hash each separately and then encode the fixed-size digests: \texttt{keccak256(abi.encode(keccak256(bytes(action)), keccak256(bytes(param))))}.
  \item \textbf{Property tests and fuzzing:} Add invariants (no two distinct tuples yield the same key) and fuzz over adversarial string/bytes splits to catch collisions in tests.
\end{itemize}

Use \texttt{abi.encode} for preimages, or include explicit separators/lengths. Prefer EIP-712 typed data for signatures to bind types and domain~\cite{chu2023survey,vidal2024vulnerability}.

\subsection{Network and Protocol Layer Vulnerabilities}
\paragraph{Beyond individual contracts, cross-domain protocols such as bridges, relayers, and signature schemes introduce systemic risks when message semantics are not cryptographically bound to their origin. This family captures replay, double-mint, and finality-mismatch vulnerabilities that arise from missing domain separation, nonces, or confirmation depth.}

\subsubsection{Replay Attack}\label{vuln:replay}
\paragraph{\textbf{Definition:}}  
A \emph{replay attack} occurs when a valid transaction or signed message intended for one context (e.g., a particular chain, domain, or session) is maliciously or inadvertently re-submitted in a different context where it has undesired effects. In smart-contract systems this commonly manifests as (a) signature replay across chains or domains, and (b) re-execution of messaging/bridge instructions due to missing nonces or domain separation~\cite{jiao2024survey, kezadri2025ethereum, atzei2017survey}.

\begin{lstlisting}[caption={Replay Attack Vulnerability}, label={lst:replay_vuln}]
pragma solidity ^0.6.0;
contract SimpleRelayer {
    address public owner;
    constructor(address _owner) public { owner = _owner; }
    // Vulnerable: verifies a signature without domain/nonce/expiry
    function execute(bytes32 payload, bytes memory sig) public {
        // msgHash = keccak256(payload)  -- no chainId, no nonce, no deadline
        // <yes> <report> REPLAY_ATTACK
        bytes32 msgHash = keccak256(abi.encodePacked(payload)); // (Line 8) missing domain/nonce/expiry
        address signer = recoverSigner(msgHash, sig);           // (Line 9) recovers over weak digest
        require(signer == owner, "not authorized");             // (Line 10) no replay guard or scoping
        // perform privileged action tied only to payload
        _doAction(payload);                                     // (Line 11) effect re-executable elsewhere
    }
    function _doAction(bytes32) internal {
        // privileged behavior...
    }
    function recoverSigner(bytes32 h, bytes memory sig) internal pure returns (address) {
        // standard ecrecover wrapper (omitted)
    }
}
\end{lstlisting}

\paragraph{\textbf{Explanation of Vulnerability:}}
\begin{itemize}
  \item \textbf{Lines 8 — Digest lacks domain separation } The signed hash is computed over \texttt{payload} alone via \texttt{keccak256(abi.encodePacked(payload))}, omitting chain id, contract address, and purpose. Identical bytes validate across contracts/chains.
  \item \textbf{Line 8 — No nonce / single-use token } The digest contains no per-user or per-message nonce, so the same signature can be submitted repeatedly (same chain) or at a later time.
  \item \textbf{Line 9 — Authorization over weak message } \texttt{recoverSigner} authenticates the signer but not the \emph{context} of execution; recovering over an ambiguous digest makes cross-domain reuse trivial.
  \item \textbf{Line 10 — Missing replay guard } No \texttt{used[signer][nonce]} or \texttt{consumed[msgHash]} mapping is checked/updated; the message remains valid indefinitely.
  \item \textbf{Line 11 — Privileged effect tied only to payload } The privileged action binds to \texttt{payload} without any consumed-ticket semantics, enabling the same logical instruction to be re-applied.
\end{itemize}

\paragraph{\textbf{Why This is a Vulnerability:}}
\begin{itemize}
  \item \textbf{Cross-context validity:} A signature over a bare payload is valid wherever identical bytes are accepted (other chains, mirror contracts, replay via relayers), so intent cannot be constrained.
  \item \textbf{Idempotence / re-use:} Without single-use nonces or deadlines, attackers can replay the same authorization to repeat the effect.
  \item \textbf{Bridge and cross-domain risk:} Relayers/bridges that omit replay filters allow previously processed instructions to be executed again on the destination chain~\cite{jiao2024survey, kezadri2025ethereum}.
  \item \textbf{High impact:} Leads to double-spends, duplicate withdrawals/mints, or replay of administrative operations.
\end{itemize}

\paragraph{\textbf{Exploitation Scenario:}}
\begin{itemize}
  \item \textbf{Cross-chain signature replay:} The owner signs ``withdraw 100'' intended for Chain~A. An adversary replays the same bytes against an equivalent contract on Chain~B; recovery succeeds and funds are unjustly withdrawn on B.
  \item \textbf{Same-chain multi-use replay:} The attacker resubmits the same signed payload multiple times to repeat a privileged action because no nonce was consumed on first execution.
  \item \textbf{Bridge message re-execution:} A previously processed bridge message is re-submitted to the destination contract; lacking consumed-message tracking, the contract mints a second time.
\end{itemize}

\paragraph{\textbf{Mitigation Recommendations:}}
\begin{itemize}
  \item \textbf{Domain separation (EIP-712 / EIP-191):} Sign structured data bound to a domain (name, version, \texttt{chainId}, \texttt{verifyingContract}). Verify on-chain against the same domain.
\begin{lstlisting}[language=Solidity]
// Pseudocode: EIP-712 digest = keccak256(
//   "\x19\x01", DOMAIN_SEPARATOR, keccak256(encodeTyped(struct)))
\end{lstlisting}
  \item \textbf{Nonces and single-use tokens:} Include a per-user/per-message nonce and record consumption.
\begin{lstlisting}[language=Solidity]
mapping(address => mapping(uint256 => bool)) public usedNonce;
require(!usedNonce[signer][nonce], "replay");
usedNonce[signer][nonce] = true;
\end{lstlisting}
  \item \textbf{Expiration windows:} Require \texttt{deadline >= block.timestamp} in the signed data; reject stale messages.
  \item \textbf{Chain-id binding:} Include \texttt{chainId} in the signed content (via EIP-712 domain or explicit field) so signatures are invalid on other chains.
  \item \textbf{Bridge idempotency:} Track and reject re-processing of \texttt{messageId}/\texttt{txHash} on destination chains; persist proofs/nonces to enforce one-time execution.
  \item \textbf{Hardened libraries:} Use well-audited EIP-712 helpers and signature-verification libraries that enforce domain, nonce, and deadline checks by default.
  \item \textbf{Testing \& analysis:} Add tests for cross-chain/same-chain replay; use static/dynamic analyzers to flag verification code lacking domain, nonce, or expiry~\cite{jiao2024survey, kezadri2025ethereum}.
\end{itemize}

\paragraph{\textbf{Notes:}}
Replay protection is foundational for cross-domain security; recent surveys and bridge incident analyses emphasize domain separation, nonces, deadlines, and explicit message-IDs as standard defenses against replay attacks~\cite{jiao2024survey, kezadri2025ethereum, atzei2017survey}.
\subsubsection{Cross-Chain Finality/Reorg Exploit (bridge double-mint)}\label{vuln:finality-reorg}
\paragraph{\textbf{Definition:}}
If a bridge honors source-chain events before they are final (or with weak/optimistic proofs), a later reorg can remove the source \texttt{Lock} while the destination \texttt{Mint} remains, producing unbacked assets (double-mint)~\cite{jiao2024survey, kezadri2025ethereum}.

\begin{lstlisting}[caption={Cross-Chain Finality/Reorg Exploit Vulnerability},label={lst:bridge-finality}]
/*
  Pseudocode (destination chain):
*/
contract LazyBridge {
    mapping(bytes32 => bool) public processed;

    // Vulnerable: accepts "lock" events with weak/optimistic proof and shallow confs
    function redeem(bytes32 eventId, bytes calldata lightProof, address to, uint256 amt) external {
        require(verifyLightProof(lightProof, eventId), "invalid proof");        // (Line 9) weak proof acceptance
        require(!processed[eventId], "already processed");                      // (Line 10) replay guard local only
        processed[eventId] = true;
        _mint(to, amt);   // (Line 12) Vulnerability: mint before deep finality/dispute window
    }

    function verifyLightProof(bytes calldata, bytes32) internal pure returns (bool) {
        // returns true for brevity; real impl would check validator sigs, etc. // (Line 16) trivial proof stub
        return true;
    }

    function _mint(address, uint256) internal { /* mint on destination */ }
}
\end{lstlisting}

\paragraph{\textbf{Explanation of Vulnerability:}}
\begin{itemize}
  \item \textbf{Lines 8–12 — Mint on optimistic proof } \texttt{redeem()} accepts a light/optimistic proof and immediately mints, without waiting for deep confirmations, finalized headers, or a dispute window.
  \item \textbf{Line 10 — Local-only replay guard } \texttt{processed[eventId]} prevents re-processing on the destination chain but does \emph{not} tie the mint’s validity to the source event’s persistence.
  \item \textbf{Line 16 — Insufficient verification } \texttt{verifyLightProof} is a placeholder for shallow checks; in practice, weak committee signatures or short confirmations are not robust against reorganizations or equivocation.
  \item \textbf{Protocol boundary:} Asynchronous consensus across chains means history can diverge temporarily; without finality-aware proofs, a destination mint may outlive a pruned source lock.
\end{itemize}

\paragraph{\textbf{Why This is a Vulnerability:}}
\begin{itemize}
  \item \textbf{Finality mismatch:} Destination state is updated (mint) on evidence that can be revoked on the source (reorg), creating unbacked liabilities.
  \item \textbf{Asymmetric reversibility:} Source \texttt{Lock} can disappear after a reorg, while destination \texttt{Mint} is irreversible (no automatic unwind), leading to permanent supply inflation.
  \item \textbf{Weak trust base:} Light/optimistic proofs with shallow confirmations or small validator quorums are vulnerable to short-range reorgs and byzantine signer sets.
  \item \textbf{High impact domain:} Bridges custody or synthesize large asset values; a single exploit yields systemic losses (double-mint, under-collateralization, insolvency).
\end{itemize}

\paragraph{\textbf{Exploitation Scenario:}}
\begin{itemize}
  \item \textbf{Timed lock + instant redeem:} The attacker locks on Chain~A at shallow depth, then immediately submits a light proof to Chain~B and receives a mint.
  \item \textbf{Induced reorg / equivocation:} The attacker (or colluding proposers) triggers a short-range reorg on Chain~A that removes the lock transaction. Chain~B retains the minted tokens.
  \item \textbf{Exit with profit:} The attacker trades/bridges out the unbacked tokens on Chain~B before any manual reconciliation, realizing profit while the system is under-collateralized.
\end{itemize}

\paragraph{\textbf{Mitigation Recommendations:}}
\begin{itemize}
  \item \textbf{Finality-aware proofs:} Require finalized headers (e.g., Byzantine Fault Tolerance (BFT) finality, checkpointed epochs) or zk/light-client verification of inclusion against finalized states before minting.
  \item \textbf{Dispute / challenge windows:} Use optimistic proofs only with a challenge period; defer mint or mint-to-escrow until the window elapses without a valid fraud proof.
  \item \textbf{Depth thresholds and reorg handling:} Enforce conservative confirmation depths on probabilistic-finality chains and implement automatic \emph{rollback or clawback} procedures if the source event is later orphaned.
  \item \textbf{Two-phase settle:} Mint to an escrowed/“pending” balance first; release to the user only after finality is attained, or allow slashable relayers to underwrite early release.
  \item \textbf{Cross-chain replay/uniqueness:} Bind \texttt{eventId} to source chain id, height, and Merkle/accumulator root; store a cryptographic commitment so that removed events cannot be proven later.
  \item \textbf{Economic security:} Require bonded relayers/guardians with slashing for incorrect attestations; size bonds relative to TVL and potential reorg depth.
  \item \textbf{Rate limits and circuit breakers:} Cap per-block/per-epoch mints and enable pausability on proof anomalies or reorg detection.
  \item \textbf{Monitoring and audits:} Continuously monitor reorg depth on source chains; formally verify proof-verification logic and finalize/rollback state machines.
\end{itemize}

\paragraph{\textbf{Practical Mapping and Tool Alignment.}}  
To bridge the conceptual taxonomy of Section~\ref{sec:taxonomy} with concrete detection practice,  
Table~\ref{tab:taxonomy} summarizes how each vulnerability family manifests at the program-analysis level.  
It lists representative \emph{detection signals} commonly used by static, dynamic, and learning-based tools,  
and indicates which major datasets or surveys include examples of each category.  
This mapping helps researchers and auditors quickly align vulnerability families with  
existing evaluation benchmarks and detection frameworks, facilitating unified reporting and tool comparison.  

\begin{table}[H]
\centering
\caption{Consolidated taxonomy of smart-contract vulnerability families with representative detection signals and datasets/tools.}
\label{tab:taxonomy}
\footnotesize
\begin{tabular}{p{3.2cm} p{4.6cm} p{4.6cm}}
\toprule
\textbf{Family} & \textbf{Detection signals (typical)} & \textbf{Datasets / Tools / Surveys} \\
\midrule
Control-Flow Interaction &
External call before state update; recursive entry; ordering/race on shared state &
SmartBugs~\cite{ferreira2020smartbugs}; SolidiFI~\cite{ghaleb2020effective}; tool surveys~\cite{he2023detection,chu2023survey} \\
External Call \& Error Handling &
Unchecked \texttt{call/send/transfer}; swallowed revert; push-payment anti-pattern &
SmartBugs~\cite{ferreira2020smartbugs}; SolidiFI~\cite{ghaleb2020effective}; surveys~\cite{he2023detection,chu2023survey} \\
State / Storage Integrity &
Untrusted \texttt{delegatecall}; proxy target drift; slot overlap; unsafe Yul writes &
Upgrade/storage analyses~\cite{iuliano2024smart,pasqua2023enhancing}; taxonomies~\cite{jiao2024survey,chu2023survey} \\
Arithmetic \& Type Safety &
Overflow/underflow (pre-0.8); div-by-zero; narrow casts &
SmartBugs~\cite{ferreira2020smartbugs}; classic surveys~\cite{atzei2017survey,ghaleb2020effective,he2023detection} \\
Environment Dependency &
Use of \texttt{block.timestamp/number/hash} for control or “randomness” &
Randomness/miner-bias studies~\cite{qian2023demystifying,sun2023assbert}; surveys~\cite{chu2023survey} \\
Access Control \& Authentication &
Missing \texttt{require}/modifiers; \texttt{tx.origin}; unprotected \texttt{selfdestruct} &
Core surveys~\cite{atzei2017survey,chu2023survey}; selfdestruct handling~\cite{colin2024integrated} \\
ABI / Input Validation &
Short-address; packed-ABI collisions; unchecked \texttt{bytes} decoding; no EIP-712 &
Packed-ABI~\cite{vidal2024vulnerability,chu2023survey}; datasets/surveys~\cite{ghaleb2020effective,he2023detection} \\
Network / Protocol Layer &
Domainless signatures; missing nonces/expiry; bridge finality/reorg and replay &
Bridge/cross-chain~\cite{jiao2024survey,kezadri2025ethereum}; classic replay~\cite{atzei2017survey} \\
\bottomrule
\end{tabular}
\end{table}

% ============================================================

\section{Discussion}\label{sec:discussion}
% ============================================================

\noindent\textbf{Consolidation of perspectives.}
Our eight-family taxonomy unifies diverse vulnerability lists and ad-hoc labels under four recurring structural dimensions: 
(i) control-flow and execution ordering, 
(ii) external-call and exception handling, 
(iii) state integrity and privilege boundaries, and 
(iv) environmental or cross-domain assumptions.  
By focusing on these first-cause axes, the framework directly connects root defects with mitigation strategies—for example, Checks–Effects–Interactions (CEI) for reentrancy and transaction-ordering flaws, or domain separation for cross-chain replay prevention.

\medskip
\noindent\textbf{Implications for detection and benchmarking.}
Instead of counting isolated bug names, tool evaluations can now report coverage along structural axes.  
This perspective exposes where static, dynamic, or ML-based detectors overlap or diverge and clarifies which analysis signals (e.g., control-flow slicing, privilege validation, or ABI decoding) each approach captures.  
Re-labeling benchmark datasets such as SmartBugs and SolidiFI according to these unified families can reduce naming drift and prevent double-counting of near-synonymous issues.  
In practice, auditors gain a clearer mapping between reported findings and root-cause categories, while tool builders obtain standardized coverage metrics.

\medskip
\noindent\textbf{Explainability and multi-axis incidents.}
Some exploits cross multiple structural boundaries—for example, ERC-20 approval races mix ordering and authorization faults.  
Following our \emph{first-cause principle}, each case is assigned to the dimension that best explains the exploit with minimal auxiliary assumptions.  
This rule improves consistency in dataset labeling and in human-interpretable tool output, where each alert can reference the triggering lines, violated principle, and a minimal counterexample trace.

\noindent\textbf{Limitations and scope.}
The taxonomy currently targets EVM-compatible ecosystems; porting to non-EVM or BFT-based chains may require adjusting environmental and cross-domain categories.  
Machine-learning models trained on legacy labels will require re-annotation to align with the unified schema—a one-time adaptation that yields more consistent evaluation and clearer coverage reporting.  
Overall, the framework provides a reusable lens for researchers, educators, and auditors to communicate about vulnerabilities with shared terminology.

% ============================================================
\section{Legacy and Historical Attack Terminology}\label{sec:legacy} 
% ============================================================

\noindent\textbf{Motivation.}
Terminology describing smart-contract vulnerabilities has evolved through incident reports, community discussions, and early static-analysis tools.  
Labels such as “gasless send,” “short address,” or “call to unknown” were often coined ad-hoc, capturing symptoms rather than structural causes.  
Although subsequent taxonomies (e.g., Decentralized Application Security Project (DAPS), SWC, and academic surveys~\cite{atzei2017survey,chu2023survey,vidal2024vulnerability,jiao2024survey,kezadri2025ethereum}) brought partial standardization, inconsistent naming still hampers comparison across tools and benchmarks.

\noindent\textbf{\\Mapping rationale.}
We align these legacy terms with our root-cause-oriented taxonomy from Section~\ref{sec:taxonomy}, following the \emph{first-cause principle}—each legacy label is mapped to the structural dimension that most directly explains its exploit mechanics. This mapping clarifies equivalences between historical and modern terminologies, enabling backward compatibility for datasets and reproducible evaluation.

\noindent\textbf{\\Mapping table.}
Table~\ref{tab:legacy-historical-alt} catalogs widely used legacy vulnerability names alongside their historical or alternative terminology found in prior research, tools, and practitioner discourse. These include both formal terms coined in academic papers and informal labels popularized through audit reports or community lore. The mapping serves two key purposes: (1) to enable interpretability and compatibility with datasets or tools that still use legacy labels, and (2) to highlight how terminology has evolved and converged toward structured, root-cause–based classifications.

\begin{table}[H]
\centering
\caption{Legacy vulnerability terms and their alternative or historical counterparts.}
\label{tab:legacy-historical-alt}
\begin{tabular}{p{6.5cm} p{7.5cm}}
\toprule
\textbf{Legacy Term} & \textbf{Alternative / Historical Term(s)} \\
\midrule
Reentrancy & Recursive Call, Reentrant Call \\
Transaction Ordering Dependence (TOD) & Front-running, Race Condition \\
Checks–Effects–Interactions Violation & Call Before State Change, Unsafe Call Order \\
Delegatecall Injection & Context Confusion, Proxy Takeover, Logic Hijack \\
Unprotected Selfdestruct & Suicidal Contract, Accidental Destruction \\
\texttt{tx.origin} Misuse & Phishing via \texttt{tx.origin}, Origin Confusion \\
Short Address Attack & Calldata Truncation Bug, ABI Padding Mismatch \\
Packed-ABI Hash Collision & Hash Collision in Commitments, Ambiguous Encoding \\
Unchecked Low-Level Calls & Ignored Return Values, Fault-oblivious Call \\
On-Chain Randomness Misuse & Miner-Manipulated Randomness, Block Entropy Exploit \\
Approval Race Condition & ERC20 Allowance Double-Spend, Approve/TransferFrom Race \\
Access Control Misconfiguration & Missing \texttt{onlyOwner}, Weak Privilege Checks \\
Bridge Replay / Finality Exploit & Double Mint, Cross-Chain Reorg Attack \\
Signature Replay Attack & Domainless Signature Reuse, Nonce-less Authorization \\
Integer Overflow/Underflow & Arithmetic Bug, Wraparound Error \\
Division by Zero & Math Error, Runtime Arithmetic Exception \\
Unchecked Assembly Decoding & Raw Memory Access, Unsafe Byte Parsing \\
DoS via Revert or Gas Exhaustion & Griefing Attack, Fallback Denial \\
\bottomrule
\end{tabular}
\end{table}

% ============================================================
\section{Conclusion}\label{sec:conclusion}
% ============================================================
This study presented an eight-family taxonomy that organizes Ethereum smart-contract vulnerabilities by their program-structural root causes rather than by inconsistent historical names.  
Each family was illustrated through representative code examples, exploit mechanics, and practical mitigations, revealing how diverse incidents converge on a small set of recurring design flaws.  
By aligning terminology, detection signals, and datasets, the taxonomy improves cross-tool comparability and fosters explainable, root-cause-oriented analysis.

\medskip
\noindent\textbf{Practical impact.}
For auditors and developers, the taxonomy serves as a diagnostic checklist that links specific bug patterns to their underlying structural causes and canonical defenses.  
For tool builders and researchers, it defines common evaluation axes, simplifying coverage measurement and dataset harmonization.  
For educators, it offers a coherent structure for teaching smart-contract security without relying on legacy, symptom-based labels.

\medskip
\noindent\textbf{Future work.\\}
Future directions include:
(i) expanding the taxonomy to cover economic-layer and oracle manipulation patterns;  
(ii) formalizing bridge and layer-2 dispute mechanisms;  
(iii) releasing a re-labeled benchmark with per-axis coverage metrics and explainability artifacts; and  
(iv) evaluating LLM-assisted detectors that output structured rationales aligned with our taxonomy.  
Collectively, these efforts aim to move the ecosystem from fragmented naming toward unified, interpretable assurance.

% ============================================================

\bibliographystyle{IEEEtran}
\bibliography{references}

\end{document}